\documentclass{aastex63}

\usepackage{hyperref}
\usepackage{natbib}
\usepackage{graphicx}

\usepackage{comment}
\usepackage{gensymb}

\usepackage{amsmath}
\usepackage{amssymb}

\shorttitle{}
\shortauthors{}

\graphicspath{{./}{figures/}}

\begin{document}

\title{The Role of Spiral Arms in Galaxies}

\author[0009-0000-3578-9134]{Bingqing Sun}
\affiliation{Department of Astronomy, University of Massachusetts Amherst, 710 North Pleasant Street, Amherst, MA 01003, USA}

\author[0000-0002-5189-8004]{Daniela Calzetti}
\affiliation{Department of Astronomy, University of Massachusetts Amherst, 710 North Pleasant Street, Amherst, MA 01003, USA}

\author[0000-0003-4569-2285]{Andrew J. Battisti}
\affiliation{Research School of Astronomy and Astrophysics, Australian National University, Cotter Road, Weston Creek, ACT 2611, Australia}
\affiliation{ARC Centre of Excellence for All Sky Astrophysics in 3 Dimensions (ASTRO 3D), Australia}

\begin{abstract}
We test the influence of spiral arms on the star formation activity of disk galaxies 
by constructing and fitting multi–wavelength SEDs for the two nearby spiral galaxies NGC 628 and NGC 4321, at a spatial scale of 1-1.5kpc scale. Recent results in the literature support the `gatherers' picture, i.e., that spiral arms gather material but do not trigger star formation. However, ambiguities in the diagnostics used to measure star formation rates (SFRs) and other quantities have hampered attempts at reaching definite conclusions. We approach this problem by utilizing the physical parameters output of the MAGPHYS fitting code, which we apply to the Ultraviolet--to--Far Infrared (UV--to--FIR) photometry, in $\geq$20 bands, of spatially--resolved regions in the two galaxies. We separate the regions in arm and interarm, and study the distributions of the specific SFRs (sSFRs=SFR/M$_{star}$), stellar ages and star formation efficiency (SFE=SFR/M$_{gas}$). We find that the distributions of these parameters in the arm regions are almost indistinguishable from those in the interarm regions, with typical differences of a factor 2 or less in the medians. These results support the `gatherer' scenario of  spiral  arms, which we plan to test with a larger sample in the near future.

\end{abstract}
\keywords{Spiral galaxies (1560) --- Spiral arms (1559) --- Star formation (1569) --- Spectral energy distribution (2129)}

\section{Introduction} \label{sec:intro}

Spiral galaxies make up more than 70\% of all massive galaxies, M$_{\textup{star}}\gtrsim$10$^{10}$~M$_{\odot}$, in the local Universe \citep[e.g.,][]{Nair+2010}. Because of the prominence of spiral arms at many wavelengths, spiral pattern formation is an important field of research and has remained for several decades, since no definite conclusion on their formation has been reached yet \citep[see recent review in][]{Sellwood+2022}.  

Irrespective of their origin and evolution, spiral patterns are important drivers of the secular evolution of galaxies, as shown by simulations  \citep[see review in][]{Kormendy+2004}. Spiral arms are most prominent at blue and mid/far--infrared wavelengths and in CO emission, suggesting that they trace recent star formation in a disk galaxy \citep{Elmegreen1981, Calzetti+2005, Koda2012, Schinnerer+2019}. The contrast between spiral arms and stellar disk  decreases at near--IR wavelengths where old stars provide the bulk of the luminosity \citep{Jarrett+2003, Meidt+2021}, but the spiral structure remains in place \citep{Buta+2010, Elmegreen+2011}. 

With this information in place, the next question is whether spiral arms act as `triggers' of  star formation or are simply `gatherers' of material. In the `triggers' scenario, when the gas enters the high density region of the spiral arm it becomes denser, possibly reaching self-gravitating conditions and/or may experience a shock  leading to star formation; this occurs because inside the corotation radius material rotates faster than the spiral pattern, while it is  slower outside of the corotation radius. In this scenario, little or no star formation occurs outside of spiral arms. In the `gatherers' scenario, spiral arms act simply as collectors of material, with little or no  impact on the process of star formation beyond providing a location for more material to be collected in the same place, which can foster star formation but does not actively induce it. A review of these scenarios is found in \citet{DobbsBaba2014}. Numerical simulations  provide support for the gatherers scenario, by predicting an increase in the SFR of a factor two or less between galaxies with and without spiral arms; in particular, these simulations find that the main effect of spiral arms is to gather gas which collects in massive  molecular clouds, which may indirectly lead to increased star formation, but the spiral arms themselves do not directly trigger the star formation \citep{Dobbs+2011, Kim+2020}. Simulations aimed at explaining the low efficiency of star formation \citep[e.g.][]{Semenov+2017} or the role of interactions in star formation \citep[e.g.][]{Tress+2020} find that star formation is regulated locally by feedback in molecular clouds and not globally by large--scale structures, supporting that spiral arms should have a secondary role. 
Many papers have tested the two scenarios observationally, reaching different conclusions. Several authors, for instance, find that the specific SFR (sSFR = SFR/M$_\textup{star}$) and the star formation efficiency (SFE = SFR/M$_{gas}$) are a factor 2--3 higher in the arms than the interarm regions, which these authors interpret as evidence in support of triggering \citep{Garcia-Burillo+1993, Seigar+2002, Martinez+2009, Rebolledo+2012, Yu+2021}. Conversely, other authors find evidence for no or only a modest excess in the arms relative to the interarm regions \citep{Elmegreen+1986, Foyle2010, Kreckel+2016, Foyle+2011, Moore+2012, Hart+2017, Querejeta2021, Querejeta+2024}. 

These discrepancies may be due to the different approaches adopted by different authors for measuring SFRs, which generally rely on the recipes
that combine UV or H$\alpha$ with IR (or use similar dust--corrected indicators). Combinations of IR and UV ($\sim$ 0.1-0.3 $\mu$m) and IR and H$\alpha$ (0.6563 $\mu$m) emission from galaxies have been calibrated and widely used as SFR indicators for close to two decades \citep{Calzetti2007, KennicuttEvans2012, Hao2011, Kennicutt+2009}.
However, those powerful diagnostics, which track both the dust obscured and unobscured star formation, have shown to have limitations when applied to regions {\em within} galaxies. For instance, the different diagnostics are sensitive to different timescales: $\sim$10~Myr for H$\alpha$ and $\sim$100~Myr for UV, while the IR can probe several different timescales, depending on the wavelength used \citep{KennicuttEvans2012}.  
As shown in \citet{Kreckel+2016}, SFR diagnostics are very sensitive to the location where the measurement is performed.  \citet{Hunt2019} demonstrates that SFR indicators can overestimate the actual SFR by factors up to 3 in low--sSFR sources.

The direct application of these diagnostics to resolved regions within a few local spirals suggests that the interarm regions host 15\%--30\% of the SFR in a galaxy \citep{Foyle2010, Querejeta2021}. 
However, the interarm UV and IR emission in the grand-design NGC 5457 (M101) shows an excess that has been interpreted as UV photons escaping from the arm regions, rather in-situ star formation \citep{Popescu2005}.
This latter interpretation is supported by resolved stellar population studies with the Hubble Space Telescope, where the interarm UV emission is found to be due to stars and star clusters formed in the spiral arms and streaming out into the interarm area, with the latter hosting little star formation \citep{Kim+2012, Crocker+2015}.
Non--ionizing UV photons and UV--bright stars are not the only components of galaxies leaking out of spiral arms and star forming regions. Ionizing photons leak out as well, contributing at
least half of the smooth, diffuse H$\alpha$ component of galaxies \citep{Thilker+2002, Oey+2007}. Both distributed, low--level star formation and shocks compete, non--exclusively, as candidates for the remaining half of the diffuse H$\alpha$ \citep{Hoopes+2003}. In addition, the diffuse component shows large variations from region to region; it is minimal in correspondence of spiral arms and regions of star formation, but represents up to 100\% in some interarm regions, thus complicating the use of H$\alpha$--based spatially--resolved SFRs. The interpretation of the IR emission from dust is even more complicated: a non--negligible fraction is heated by evolved stellar populations \citep{Draine+2007, Kirkpatrick+2014} 
and between 30\% and 80\% of the Mid--Infrared emission is diffuse and does not appear to be associated
with current or recent star formation \citep{Calzetti2007, Crocker2013, Calapa+2014}.

The amount and distribution of star formation hosted outside of spiral arms affects the interpretation of other diagnostics, including the spatially--resolved Schmidt–Kennicutt Law
\citep[SK Law, ][]{kennicutt1998}. 
The SK Law, which describes the relation between the SFR and gas surface densities in galaxies, is a key relation for connecting stars to their `fuel' and for formulating a physical model of star formation. The physical interpretation of the SK Law critically depends on the functional shape of the relation between these two parameters, and vast literature is available on this topic \citep[e.g.,][]{KennicuttEvans2012, Roychowdhury+2015}. 
Attributing the diffuse (mostly interarm) emission to either current star formation or other causes alters the interpretation of the physical mechanism(s) that produce the trend. It is still unclear whether the molecular clouds found in the interarm regions of galaxies are relatively inert shreds of the star-forming clouds found in the spiral arms \citep{Koda2012, Koda+2023} or they share many properties with the spiral arm clouds, with only subtle differences \citep{Rosolowsky+2021}. Whichever the case,  some star formation still occurs in the interarm regions \citep[e.g.][]{Foyle2010, Querejeta2021}, although existing estimates are based on the SFR diagnostics discussed above and, thus, are potentially affected by biases.

In this work, we attempt to overcome previous limitations on the derivation of SFRs by employing multi-wavelength spectral energy distribution (SED) fitting, rather than recipes that use one or two bands. Some of the fundamental properties encoded in  a population's SED include the SFR, star formation history (SFH), total stellar mass, and the physical conditions of the dust and gas 
\citep{Conroy2013ARAA}.
This approach has been shown in the literature to mitigate the issues related to recipes by leveraging of the full SED for deriving internally-consistent physical parameters \citep{Hunt2019}.

This is the first paper of a larger study aiming at investigating a sample of more than 30 nearby spiral galaxies, leveraging the large multi--wavelength archival holdings available from NED, the NASA Extragalactic Database. In this pilot study, we demonstrate our methodology on two of the sample galaxies: NGC~628 and NGC~4321. We divide each galaxy  into a few  hundreds of sightlines and construct multi–wavelength SEDs, which we attribute to spiral arms and diffuse (interarm) regions using masks from the literature. We use MAGPHYS \citep{daCunha2008MNRAS} to model each SED. We quantify the fraction of current SFR occurring in the interarm and arm regions, and compare differences in sSFR and SFE between them, both globally across each galaxy and radially after dividing each galaxy in galactocentric annuli. 

Sections \ref{sec:data} and \ref{sec:method} present the imaging data we use in this study and the processing we apply to the datasets, respectively. 
Our results for the two galaxies are presented and discussed in section \ref{sec:results}. The last section provides a brief summary of this work with its main conclusions. 

\section{Data} \label{sec:data}

We leverage the large and homogeneous surveys of nearby galaxies secured by Spitzer and Herschel \citep{Kennicutt2003SINGS, Dale2017, Kennicutt2011KINGFISH},
to construct a sample of star forming spirals. 
In addition to the imaging with the \textit{Spitzer} Infrared Array Camera (IRAC , 4 bands), the Multiband Imaging Photometer on Spitzer (MIPS, 3 bands), the \textit{Herschel}  Photodetector Array Camera and Spectrometer (PACS, 2 or 3 bands) and the Spectral Photometric Imaging Receiver (SPIRE, 3 bands), we require our galaxies to have archival data from The Galaxy Evolution Explorer (GALEX, 2 bands), optical (4+ bands, typically the 5 SDSS bands) and the Two Micron All-Sky Survey (2MASS, 3 bands) covering the entire extent of each galaxy, for a total of at least 20 bands per galaxy. The images are generally available from NED, the NASA Extragalactic Database\footnote{http://ned.ipac.caltech.edu/}, with two galaxies (NGC 3184 and NGC 6946) requiring retrieval of the GALEX images directly from MAST\footnote{The Barbara A Mikulski Archive for Space Telescopes, https://archive.stsci.edu/}, and five additional galaxies requiring retrieval of PACS and SPIRE data from the Herschel Archive\footnote{http://archives.esac.esa.int/hsa/whsa/}, with associated processing. All images are  inspected to ensure that the galaxies are detected in all bands, and can yield between several tens and several hundreds of independent lines of sight.

We impose a limit on inclination, $i\leq 60\degree$, to minimize line–of–sight confusion. We also exclude all galaxies hosting Sy1’s in their centers\footnote{ From NED, the NASA Extragalactic Database}, and mitigate the impact of the AGN in Sy2 hosts by excluding the central $\sim$1 kpc from the analysis for those galaxies where the dominance of an AGN in the central region is documented  \citep[e.g.,][]{Kennicutt2003SINGS, Moustakas+2010}. The final sample consists of 34 galaxies within $\sim$20~Mpc.

As stated in the Introduction, in this first paper we concentrate on the two sample spiral galaxies NGC 628 and NGC 4321. The two galaxies are located at distances of 9.77 and 15.20 Mpc, respectively \citep{McQuinn+2017, Freedman+2001}  and have inclinations of 5$^o$ \citep{Shostak+1984} and 27$^o$ \citep{deVaucouleurs1976}.
Due to their proximity and low inclinations, these galaxies are optimal cases for testing our approach. To resolve arm and interarm regions at 1--1.5 kpc scales, we include data only out to SPIRE250. All the images of these two galaxies are retrieved from NED, including GALEX, the Sloan Digital Sky Survey (SDSS),  2MASS, IRAC, MIPS, PACS, and SPIRE.

\section{Method}\label{sec:method}

\subsection{Image Processing}
We use images already processed by the respective facilities (e.g., GALEX SDSS, 2MASS Spitzer, Herschel), with bright foreground stars masked out to avoid contamination of photometric measurements. The masking apertures were determined by-eye, and were based on a multi-wavelength analysis of the foreground (or background) interlopers  (D. Dale, 2021, Private Communication). 

For each galaxy, we are interested in obtaining multiwavelenghth photometry of spaxels covering both arm and interarm regions.
We thus align the images of each galaxy to their respective reference frame in the Spitzer/IRAC 3.6 $\mu$m band, resamplig all images to the IRAC pixel scale of 0.75 arcsec, as described below. No astrometric corrections are applied to the images.

The images are aligned and resampled using the publicly available software SWarp \citep{Bertin1996}. The resampling process involves both filtering and interpolation between image pixels. SWarp implements oversampling on the destination grid automatically, which helps preserve image details and reduce artifacts during image transformations. As for the interpolation, we adopt the default Lanczos3 resampling which is suggested to be the best compromise in most cases. We manually set the galaxy  centers.

Subtraction of the background during resampling is important to avoid spurious effects on the resampled images, and this function can be turned on/off in the SWarp 
parameter settings. We perform the background subtraction by hand, using a uniform background level, and keep the default Swarp background subtraction off. The reason is because we use larger regions of each image than Swarp does to evaluate the background. Our choice is driven by the desire to avoid 
subtracting too much flux from the galaxies themselves, since the galaxies' emission occupies the majority area of each image or mosaic. The level of the background is determined from  visual inspection of each individual image used to build the SEDs; the results from this inspection also justify  the choice of a constant background level across each image. We confirm our choices by comparing the  total fluxes we derive from aperture photometry with those published in  \citet{Dale2017}, finding general agreement with our measurements (see below). 

The resampled images are then convolved to a common PSF, which we choose to be the PSF of our lowest resolution image for this  analysis, SPIRE 250 (FWHM$\sim$18$^{\prime\prime}$). The convolution kernels applied in this work are from \cite{Aniano2011}. Though they provide the kernel for convolving MIPS 160 to SPIRE, we exclude MIPS 160 from further consideration. It has a broader PSF, by more than a factor of 2, than SPIRE 250, thus cannot be convolved safely into the PSFs of the other instruments/bands \citep{Aniano2011}. Stars are masked with no interpolation in the images, which may cause `growth' of the masked region when convolving to the lower resolution of the SPIRE 250 PSF. For the masks of the brighter stars, that are as large as or larger than the final spaxel's size (see below), we allow the size of the region to grow by a factor $\sim$1.5 to prevent biases in the final fluxes. However, most masked regions are significantly smaller than final spaxel size; for these, we verify the integrity of the final photometry in each spaxel, confirming that the results are not affected.

Finally, all images are resampled to spaxels with sizes 18$^{\prime\prime}$ (consistent with  SPIRE 250 bands' PSF), so that each spaxel corresponds to regions $\sim$0.85 kpc and 1.33 kpc in physical size for NGC 628 and NGC 4321, respectively. The inclination--corrected areas of the spaxels are 0.73~kpc$^2$ and 1.97~kpc$^2$ for NGC 628 and NGC 4321, respectively.

\begin{figure}[hb]
  \centering
  \includegraphics[width=0.97\linewidth]{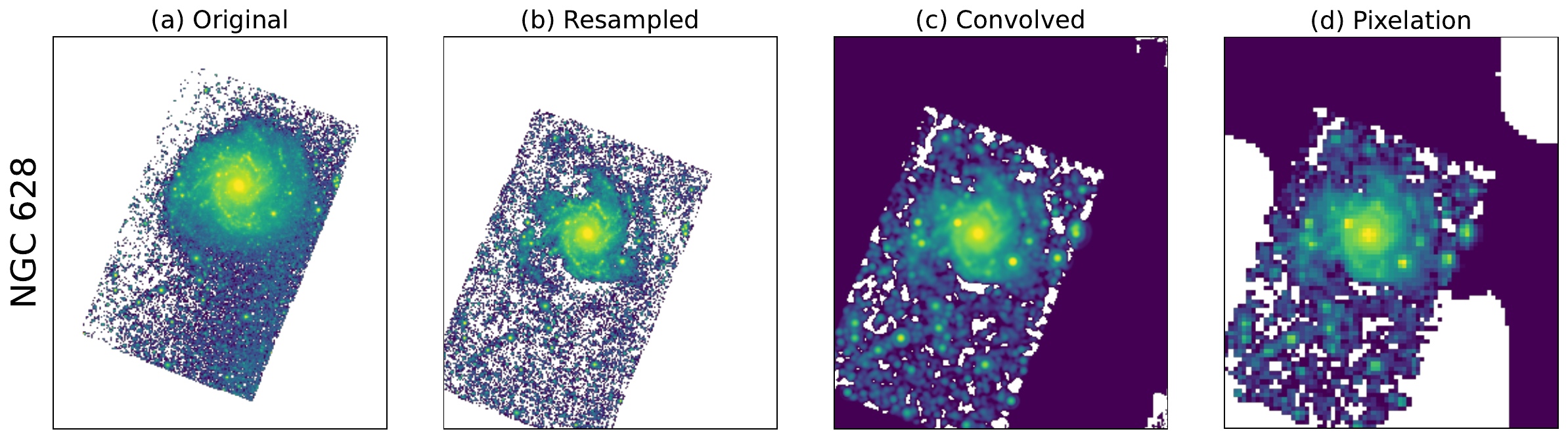}
  \caption{NGC 628 IRAC 4.5~$\mu$m image processing. Panel (a): original IRAC 4.5~$\mu$m image; (b): image projected and sampled onto the frame of the IRAC 3.6~$\mu$m image, with background  subtracted; (c): image after convolution, to match the SPIRE 250 PSF; (d): image resampled to the final spaxel scale of 18$^{\prime\prime}$. For this galaxy there are only few stars within its global aperture, so they are masked out in the last step.}
  \label{fig:NGC628processing}
\end{figure}

\begin{figure}[hb]
  \centering
  \includegraphics[width=0.97\linewidth]{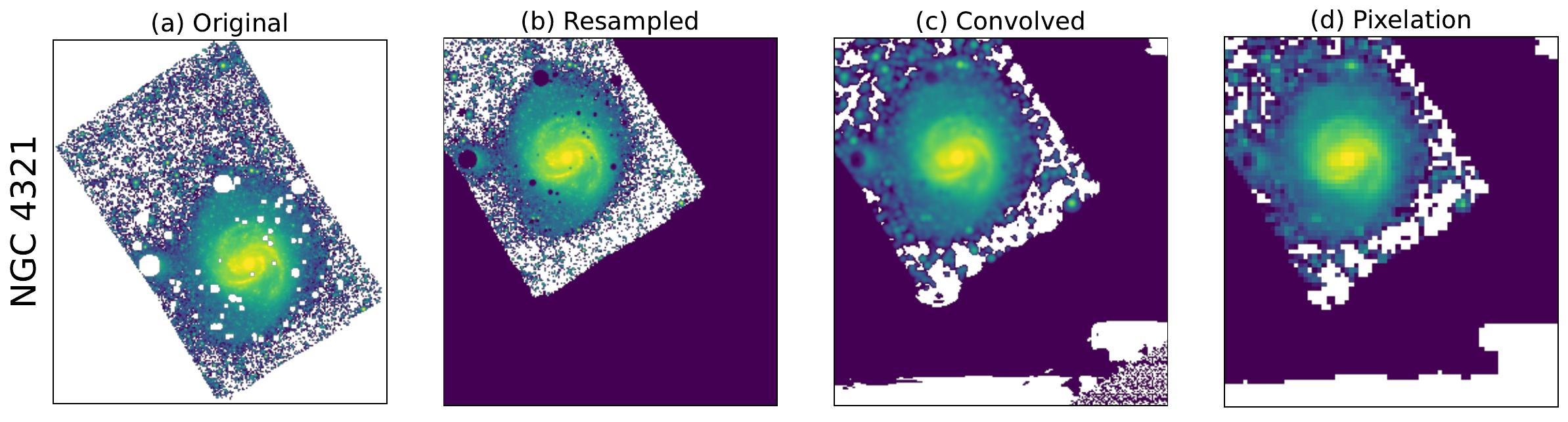}
  \caption{The same image processing as Figure~\ref{fig:NGC628processing}, but for NGC 4321. For this galaxy, stars are masked out before these steps.}
  \label{fig:NGC4321processing}
\end{figure}

We give as example the resulting  images from  the  processing of the IRAC 4.5 $\mu$m image for each of the two galaxies in figures  \ref{fig:NGC628processing} and \ref{fig:NGC4321processing}.

\subsection{Photometry}
Before constructing and fitting the SEDs of individual regions, we performed a sanity check by comparing the global fluxes measured for the two galaxies with the photometric results of \cite{Dale2017}. While we refer to their apertures for the photometry, we need to make them smaller for NGC 628 because some of our images do not cover the entire area of the galaxy. The adopted sizes and position angles are listed in Table~\ref{table:aperture}.

\begin{table}[h!]
\centering
\begin{tabular}{ cccccc } 
\hline
Name & RA (deg)  &  Dec (deg) & 2a($^{\prime\prime}$) & 2b($^{\prime\prime}$)  & PA ($\degree$) \\
\hline
NGC 4321 & 185.728463 & 15.821818, & 558.0 & 483.0 & 40.0\\
NGC 628 & 24.173946 & 15.783662 & 586.0 & 538.667 & 90.0\\
\hline
\end{tabular}
\caption{Apertures and position angles adopted for the global photometry. The RA and Dec of the apertures' centers are from \cite{Evans+2010}. }
\label{table:aperture}
\end{table}

\begin{figure}[!h]
  \centering
  \includegraphics[width=8.5cm]{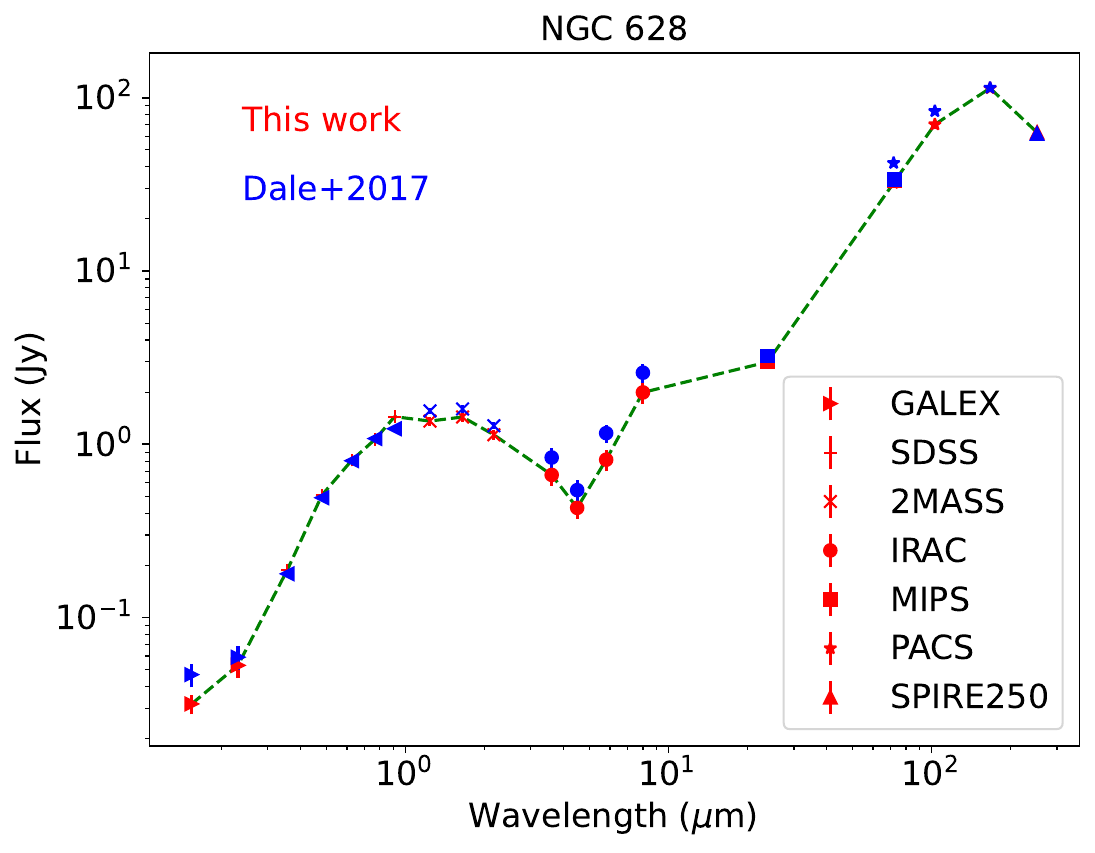}
  \includegraphics[width=8.5cm]{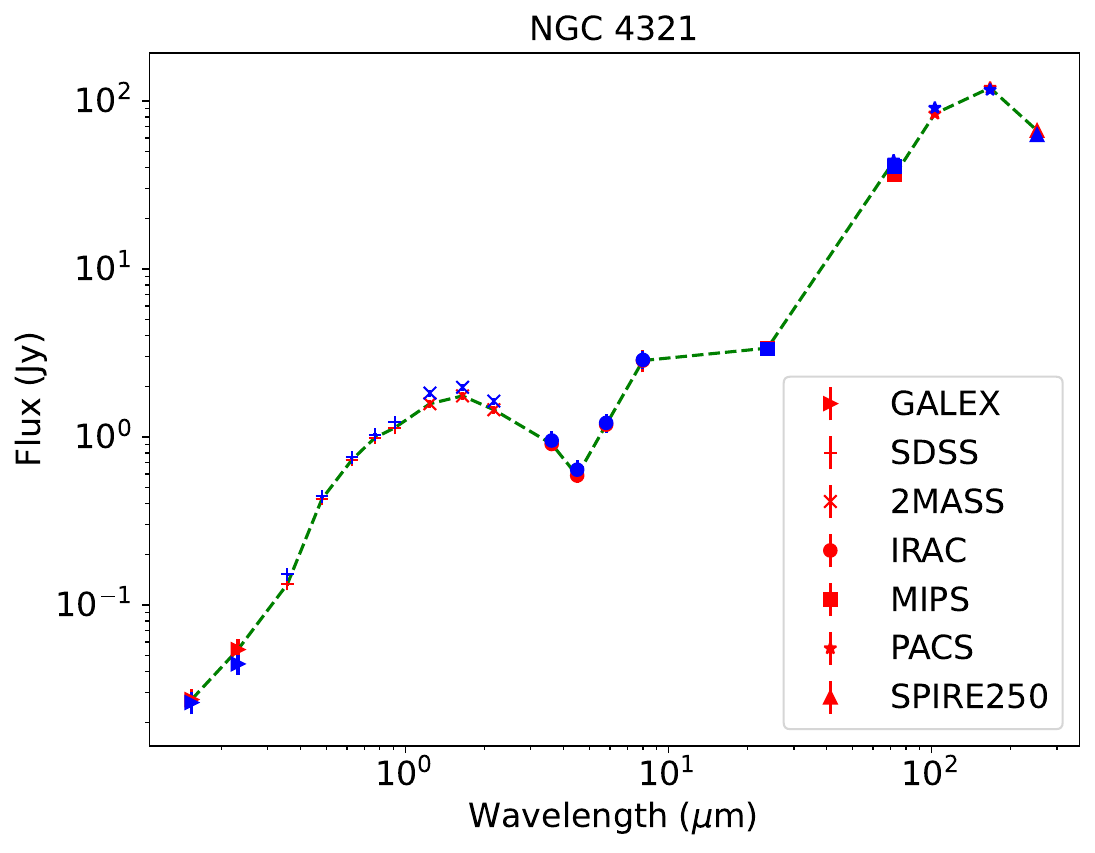}
  \caption{Left: Global flux SED for NGC 628. Different symbols represents different instruments (see the legend). Red symbols indicate the measurements in this work, and the blue symbols correspond to the measurements in \cite{Dale2017}. The green dashed line  connects our flux points to show a rough shape of the galaxy's SED.  Right: Same as the left panel but for the galaxy NGC 4321.  Error bars are calculated following \cite{Dale+2012, Dale2017}.}
  \label{fig:globalSEDs}
\end{figure}

Figure~\ref{fig:globalSEDs}  
shows our photometric measurements compared with those of \citet{Dale2017} for both galaxies and all bands. There is generally $<10\%$ discrepancy in photometry between this work and \cite{Dale2017} for NGC 628, except for the IRAC fluxes which are discrepant by more than $10\%$, but less than $20\%$. 
The nature of this discrepancy is unclear, although we cannot exclude the possibility of a small over--subtraction in the background for the IRAC images. However, we consider the discrepancy small enough that we still accept the current results. 
All bands are within $5\%$ for NGC 4321 except GALEX NUV, which are discrepant by $<10\%$. In summary,  the global fluxes we measure  for both galaxies are generally consistent with those in  \citet{Dale2017}.

\begin{figure}[!tbp]
  \centering
  \includegraphics[width=0.95\linewidth]{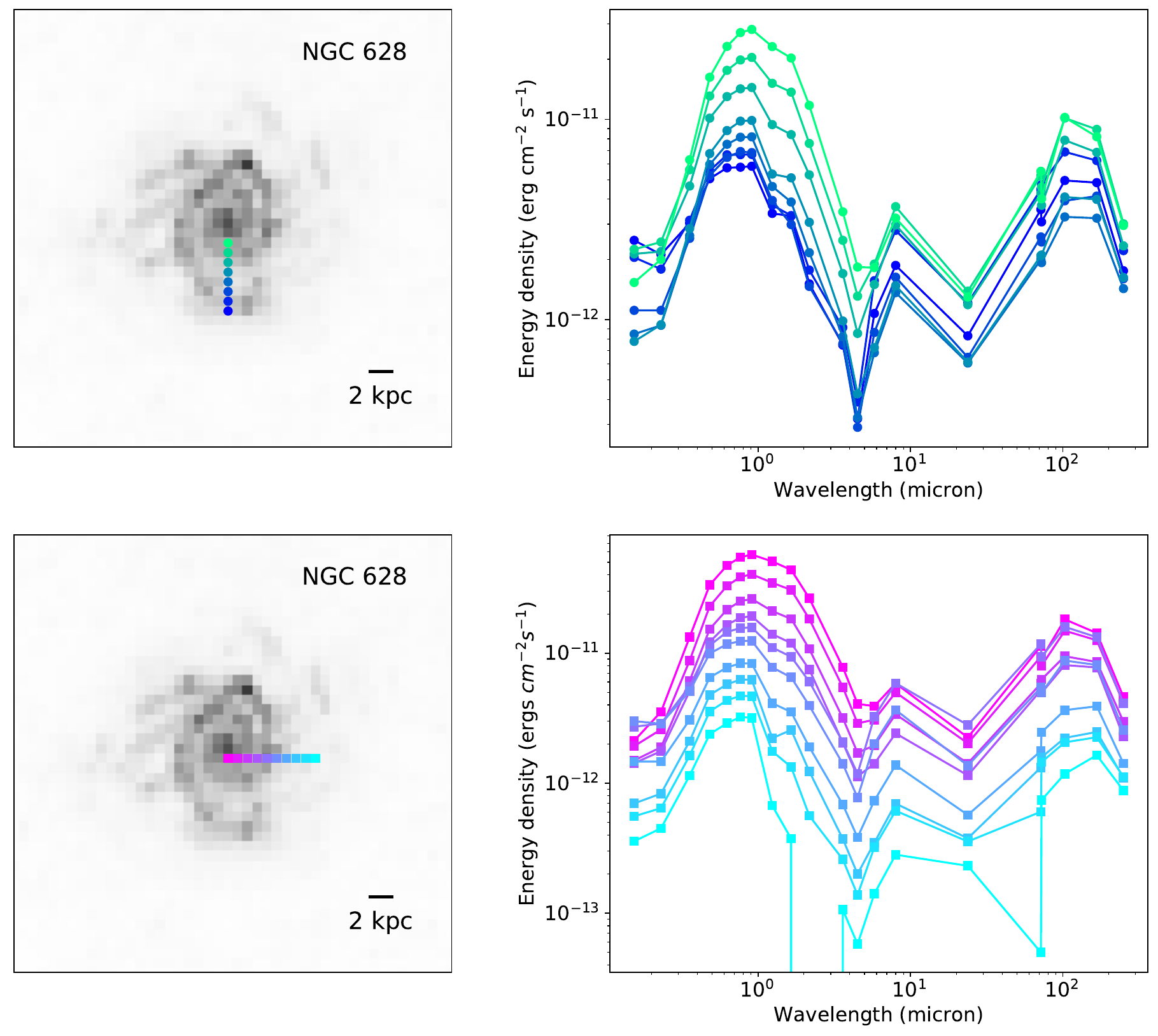}
  \caption{Examples of spaxel SEDs for NGC 628. Left panel in each row: the location of the spaxels on the SPIRE 250 image along a galactocentric radius are shown with color symbols. The spatial scale of the images is given in in the lower right corner of each panel. Right panel in each row: spaxel SEDs along this radius; the color matches the color coding of the spaxels in the Left panel.}
  \label{fig:628SEDs2Radii}
\end{figure}

\begin{figure}[!tbp]
  \centering
  \includegraphics[width=0.95\linewidth]{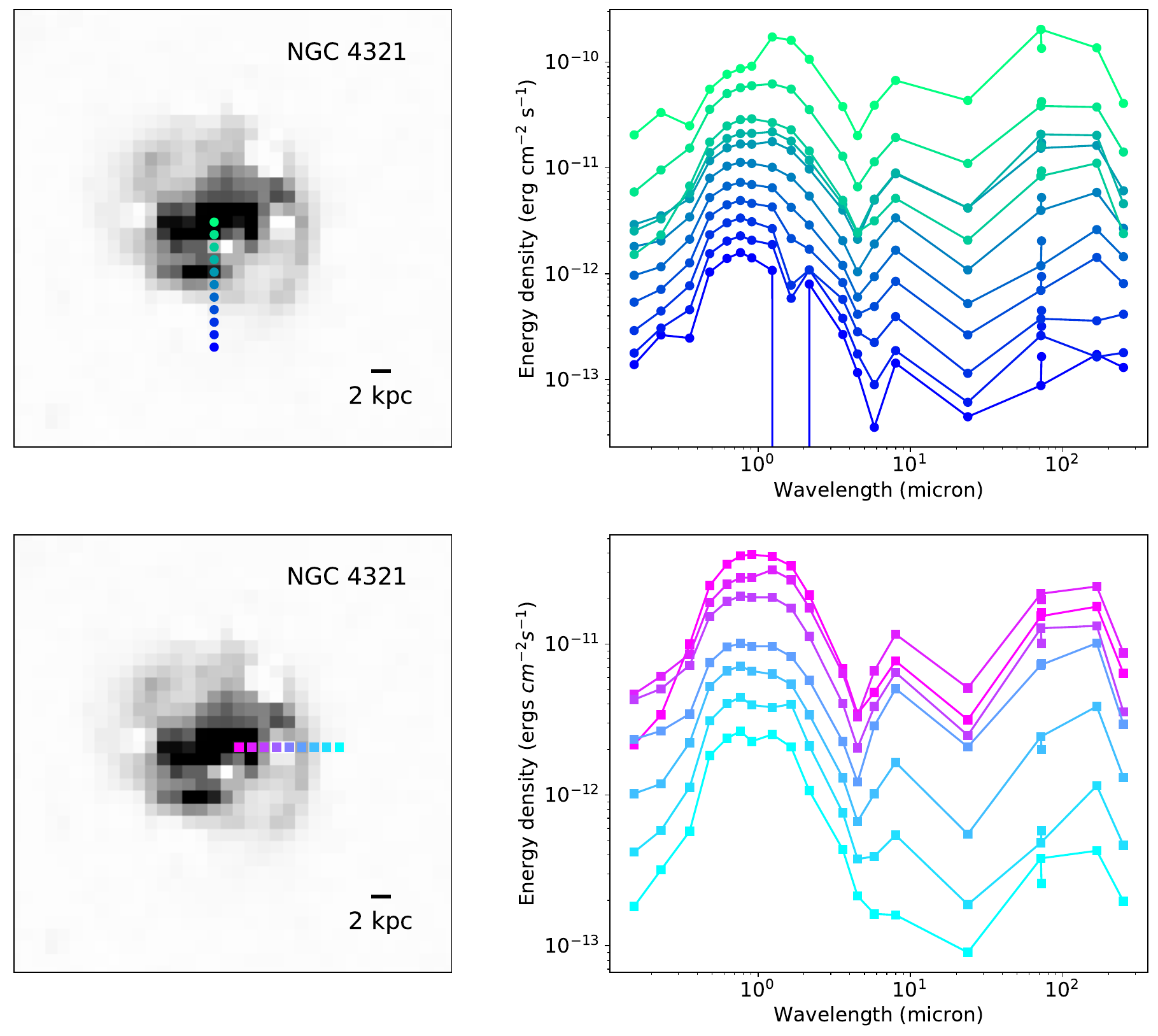}
  \caption{Same as Figure~\ref{fig:628SEDs2Radii} for NGC 4321. 
  }
  \label{fig:4321SEDs2Radii}
\end{figure}

We use the signal-to-noise ratio of the 250 $\mu$m band as our reference and we elect to only fit the SEDs of spaxels with S/N$\geq$3 in this channel, in order to ensure that enough reliable data are available to fit the dust SED. 
Figure~\ref{fig:628SEDs2Radii} and figure~\ref{fig:4321SEDs2Radii} give some examples of spaxel SEDs for NGC 628 and NGC 4321, respectively, and their relative locations in the galaxy are shown on the SPIRE 250 images.

We observe discrepancies between MIPS 70 and PACS 70 flux measurements, which is well known \citep[e.g.,][]{Aniano2012, Aniano2020}. This  discrepancy has remained mostly unsolved, and has been attributed to different treatment of the non--linear response of the two cameras.  \cite{Aniano2012, Aniano2020} suggest to include both photometric values for a better estimate of the dust parameters. 

The uncertainty on the flux measurements is  calculated in the following way.  
After plotting the spaxels' value distribution of the whole image at each wavelength, we fit a Gaussian curve to the left side of this distribution (the background). The standard deviation of the Gaussian curve is set to be the uncertainty for all spaxels in this image. As we are only including well--detected spaxels in our analysis, the error on the background subtraction is our main source of relative uncertainty. 

\subsection{Identification of the Spiral Arms}
\begin{figure}[!h]
    \centering
    \includegraphics[width=8cm]{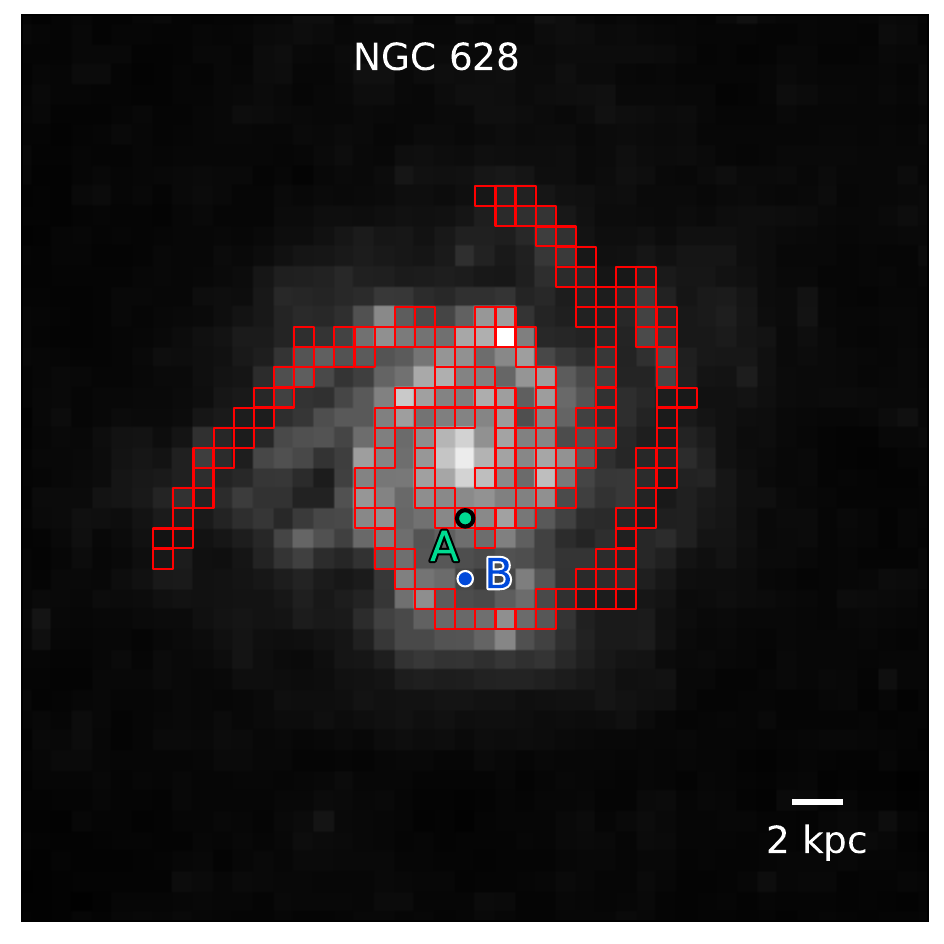}
    \includegraphics[width=8cm]{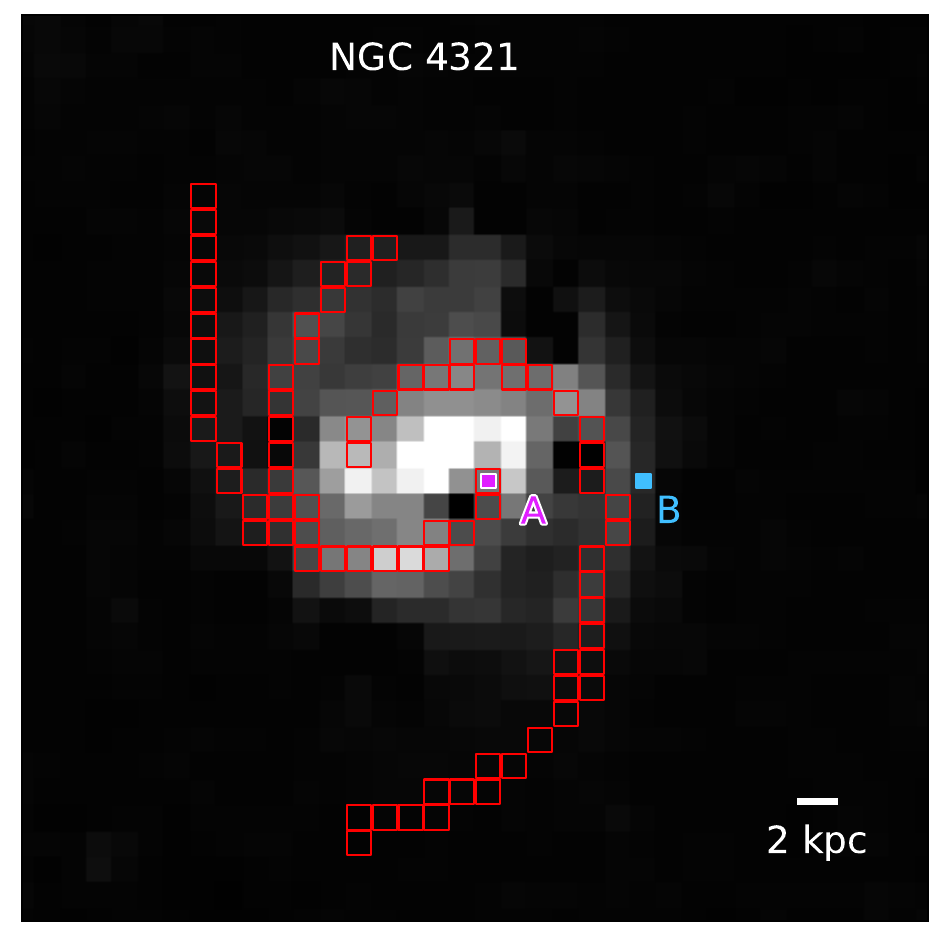} 
    \caption{Spiral arm spaxels identified in this work. 
    Left: NGC 628.  Right: NGC 4321. Both panels use red unfilled squares symbols to identify the spiral arm spaxels, overlayed on the galaxies' respective SPIRE 250 images. Spaxels marked `A' and `B' in both panels identify the spaxels we use as examples of SED fits in Fig. \ref{fig:628SEDfits} and \ref{fig:4321SEDfits}.}
    \label{fig:spiralarms}
\end{figure}

There are different approaches for identifying spiral arms in galaxies, which include dust lane images, 8 $\mu$m (dust emission) images, and other infrared images \citep[e.g.,][]{Elmegreen+1984, Kendall+2008,Foyle2010, Shabani2018,Abdeen+2020}.

Here we adopt the morphological masks recently released by \cite{Querejeta2021}, which are based on the \textit{Spitzer} IRAC 3.6 $\mu$m images. Emission in this band mainly tracks the old stellar populations, which dominates the total stellar mass. \cite{Querejeta2021} identified five environments including centers, bars, spiral arms, interarm regions and discs without strong spirals in the PHANGS (Physics at High Angular resolution in Nearby GalaxieS) sample of 74 nearby galaxies. Their sample includes both NGC 628 and NGC 4321. For more details of how the authors  define each component, we refer the reader to the method section in \cite{Querejeta2021}.

After overlaying the \cite{Querejeta2021}'s masks on our images, spaxels are marked as arm regions if more than 50\% of their area fall within the arms as defined by the masks. 
We exclude the central $\sim$1~kpc region from further consideration in our analysis, to remove the bulge region of each galaxy. Regions outside the spiral masks and outside the inner 1~kpc are considered interarm regions for the purpose of our study. 
For NGC 4321, we also exclude all spaxels associated with the central bar \citep[e.g.][]{Maeda+2023}. \cite{Querejeta2021} further define a component around the bar of this galaxy called ``interbar'', finding  that the distinction between interarm and interbar may not be significant. In our analysis, we retain this component for better statistics, which we mainly assign to the interarm spaxels.

In Appendix~\ref{AppendixB} we test two additional definitions of spaxels belonging to spiral arms: that more than 65\% and 80\%, respectively, of a spaxel's area needs to fall within the arms as defined by the masks. We show in the Appendix that these different choices  have negligible influence on our results. 

\subsection{SED Fitting}\label{sec:sedfit}

Use of SED–fitting tools to model the stellar populations of galaxies dates back at least two decades, and has consistently improved over time thanks to progress both with the modeling of stellar populations and dust emission and with the availability of homogeneous multi–wavelength archival holdings. 

One approach to model the emission from stars and dust consistently is to solve the radiative transfer equation assuming an idealised spatial distribution of stars and dust.  
However the complexity of the computations makes it hard to derive statistical constrains from large samples of galaxies.
In parallel, simple empirically calibrated infrared libraries have been developed to interpret emission from galaxies at wavelengths from 3 to 1000 $\mu$m, such as the single parameter libraries presented by \cite{CharyElbaz2001} and \cite{DaleHelou2002}. These libraries are easy to apply to large sample of galaxies,
but have the disadvantage of not directly relating the infrared dust emission to the dust absorption in stellar populations.

Several physically-motivated multi-band galaxy SED-fitting models have been developed relatively recently, all of them using as basic assumption the energy balance between the stellar radiation absorbed by dust in the UV/optical/NIR and the radiation emitted by dust in the MIR/FIR and longer wavelengths.
Currently some of the most commonly used SED-fitting codes include CIGALE  \citep[Code Investigating GALaxy Emission, ][]{Noll2009,Boquien2016,Boquien2019}, GRASIL  \citep[the GRAphite-SILicate approach, ][]{Silva1998}, MAGPHYS \citep[Multi-wavelength Analysis of Galaxy Physical Properties, ][]{daCunha2008MNRAS}, Bagpipes \citep{Carnall+2018} and Prospector \citep{Leja+2017}. 

We elect to use MAGPHYS for the following reasons: (1) MAGPHYS’s stellar population libraries cover a wide range in metallicities, between 1/50th and twice solar, which fully encompasses the range of metallicities of our sample. (2) The SFHs implemented in MAGPHYS are random bursts superimposed on exponentially decreasing SFRs, which are reasonable representations of the SFHs expected in disk galaxies \citep{Maraston+2010}. (3) The code returns, among several parameters, M$_\textup{star}$, SFRs averaged over several timescales, which we use to derive sSFRs over those timescales for each region, as well as the age of the luminosity--weighted oldest stellar population in the region. One limitation of MAGPHYS in its current distribution is the absence of nebular emission in the models; we expect negligible impact from this, as our SEDs are observed in broad bands, to which the contribution of nebular emission is minimal.

As already stated above, the basic assumption of MAGPHYS is energy conservation. \citet{Smith+2018} tested this assumption against simulations, finding that it remains valid down to $\sim$1~kpc scale. 
The stellar component in MAGPHYS is computed using the \cite{BC2003} stellar population synthesis code in the wavelength range 91 {\AA} to 160 $\mu$m, and the attenuation by dust is computed using a two-phase dust attenuation model \citep{CharlotFall2000}. 
The sum of the IR luminosity contribution by dust in the birth clouds and in the diffuse ISM is the total luminosity that is absorbed and reradiated by dust. 

The shape of the  ambient ISM SEDs at MIR wavelengths is fixed, to reduce the number of adjustable parameters in MAGPHYS. This is an acceptable approximation because previous studies have shown that variations in the average radiation field of the ambient ISM do not change the overall SED shape significantly \citep{DrainLi2007}.

The input parameters to MAGPHYS consist of the galaxy's redshift, the flux measurements and uncertainties in the different bands, as well as the filter transmission curves of the bands used, with several common ones (GALEX, SDSS, Spitzer, Herschel) already codified in the program. 
MAGPHYS performs the SED fits  in two steps. First, it assembles a comprehensive model library at the same redshift as the observed galaxy (in this work, regions).  
Second, it compares the observed SED with all the model SEDs in the library, and then builds the marginalized likelihood distributions (PDFs) of each physical parameter.  

MAGPHYS returns several physical parameters at the same time, including stellar mass, SFR averaged over the most recent 10 and 100~Myr, and the age of the luminosity--weighted oldest stellar population. 
As for many  SED fitting codes, the reliability of the results depends on several factors, including the number of data points (free parameters) and the magnitude of the measurement uncertainties. \citet{Smith+2018} used simulations to verify that the physical parameters returned by MAGPHYS for galaxies match the true parameters reasonably well down to 1~kpc scale, when using, like in our case, 21 bands from the FUV to the FIR.

\begin{figure}[!tbp]
  \centering
  \includegraphics[width=1\linewidth]{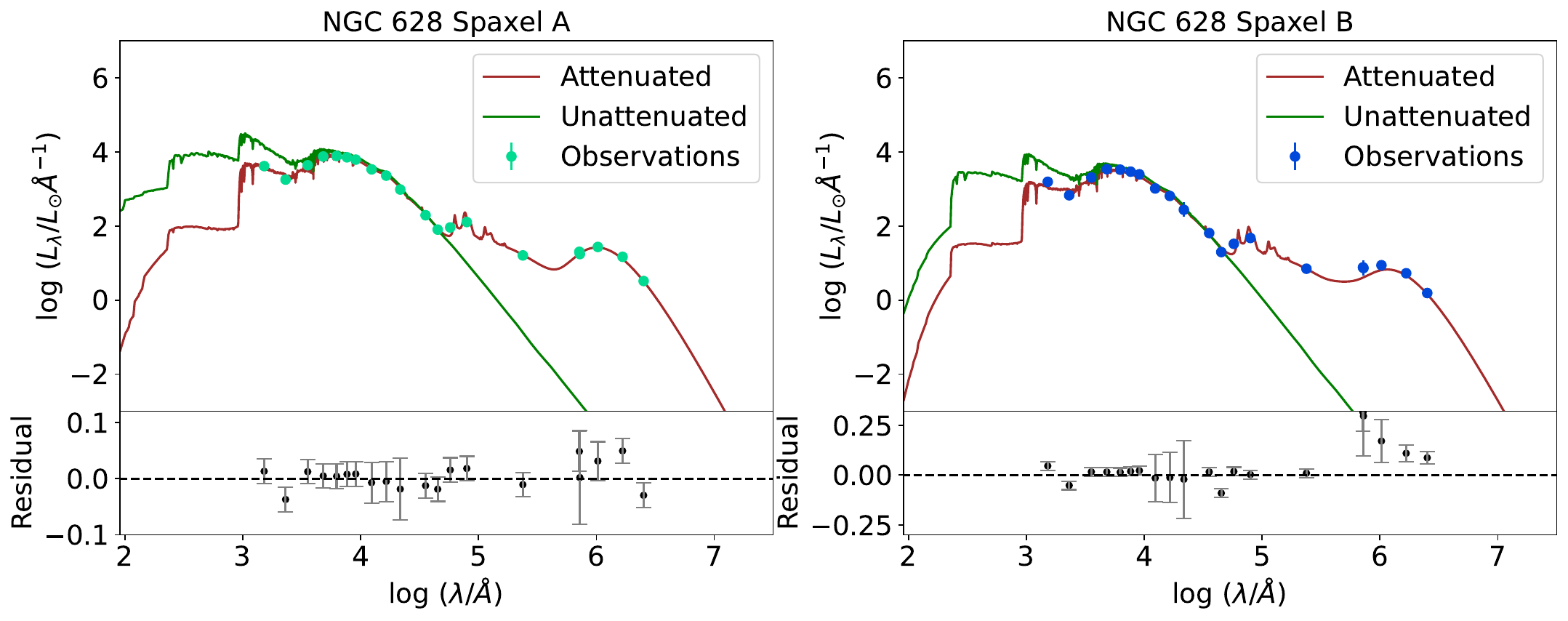}
  \caption{Best-fitting models for two spaxels in NGC 628, identified as `A' and `B' in Figure~\ref{fig:spiralarms}. Left panel: an arm region close to the center. The green points are observed fluxes. Right panel: an interarm region close to the edge of the galaxy. In both panels, the dark red line shows the best-fit model SED with dust attenuation. The fit residuals are shown in the small panels below the SEDs; note that the right panel's residuals have a different scale from the left panel. The green lines are the best-fit stellar SEDs without dust attenuation.}
  \label{fig:628SEDfits}
\end{figure}

\begin{figure}[!tbp]
  \centering
  \includegraphics[width=1\linewidth]{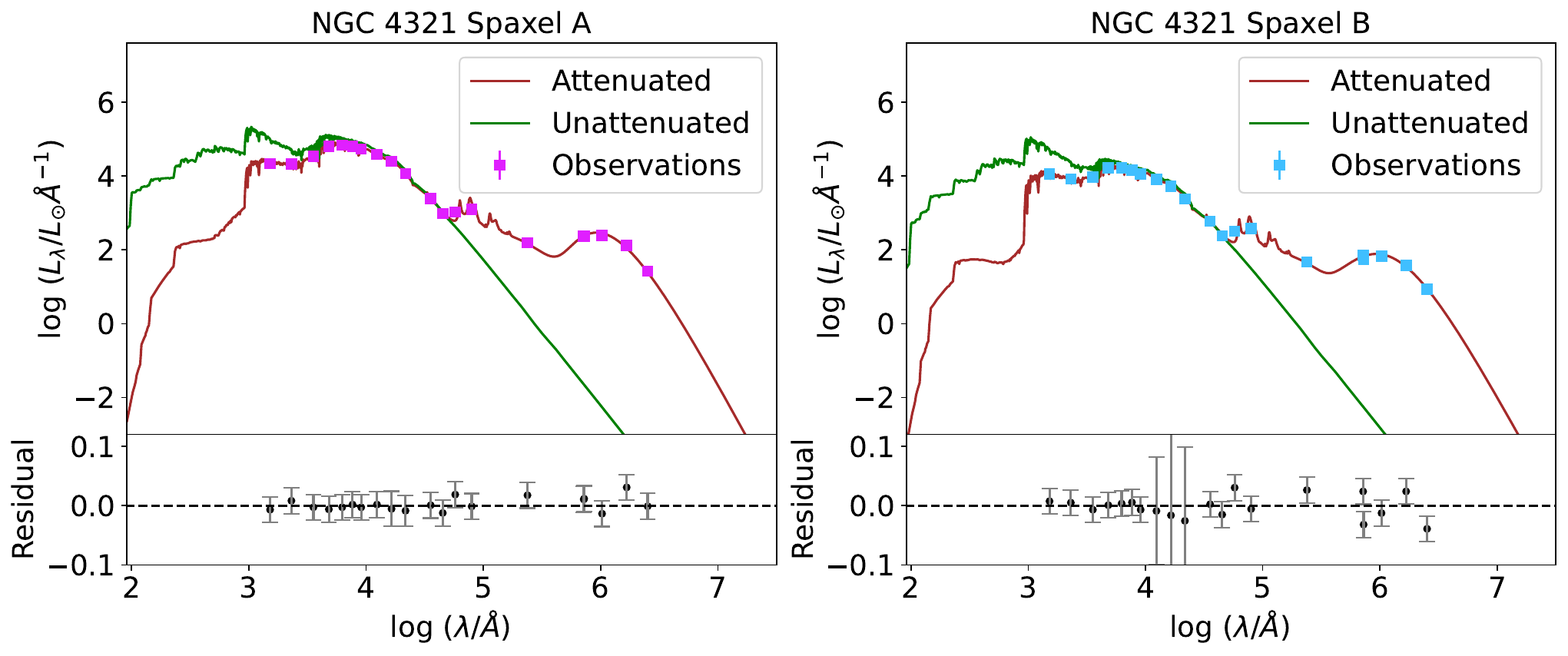}
  \caption{The same as Figure~\ref{fig:628SEDfits}, but for NGC~4321. 
  }
  \label{fig:4321SEDfits}
\end{figure}

Figures \ref{fig:628SEDfits} and \ref{fig:4321SEDfits} show some examples of the SED fits from MAGPHYS; we report the best-fit stellar SEDs, with and without dust attenuation (dark red and green lines, respectively), and the residuals of the best fit attenuated SED.

\section{Results and discussion} \label{sec:results}

\subsection{Parameter Maps}
\begin{figure}[!h]
    \centering
    \includegraphics[width= 1\linewidth]{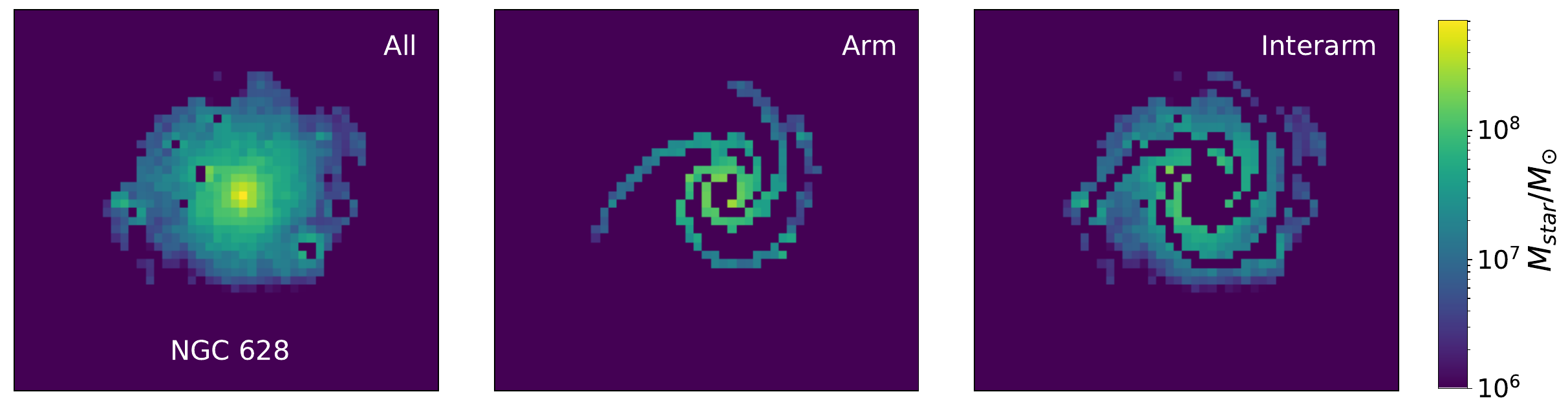} 
    \includegraphics[width=1\linewidth]{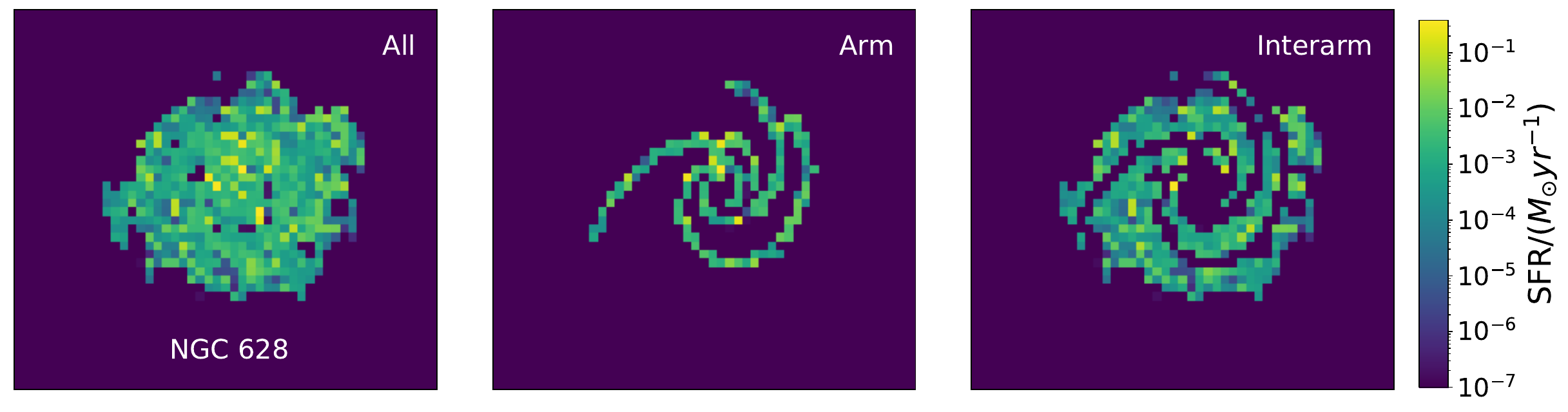}
    \includegraphics[width=1\linewidth]{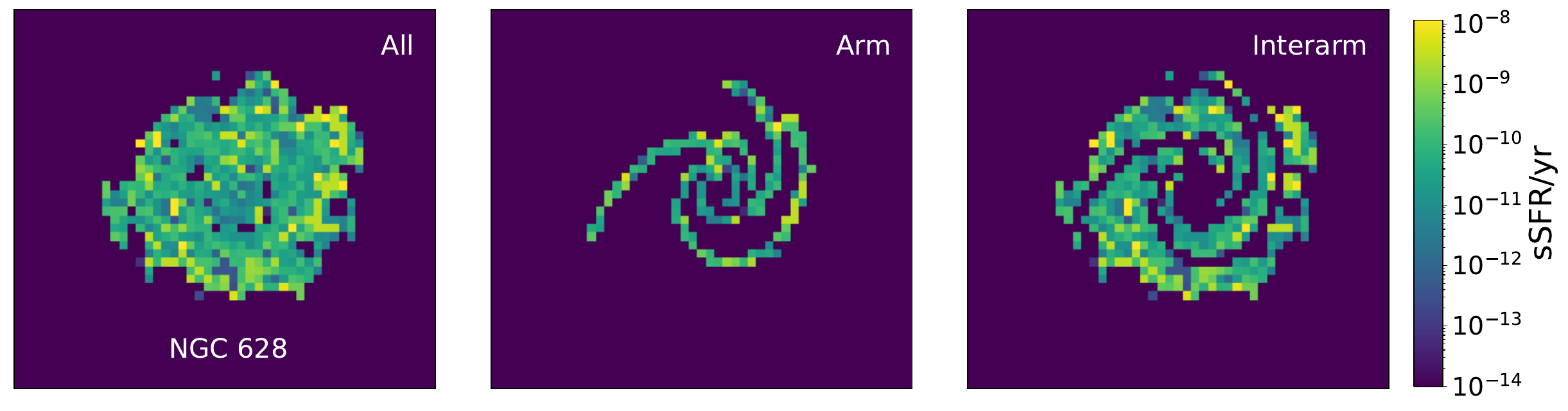}
    \caption{Parameter maps for NGC 628. Upper row: stellar mass; middle row: SFR; and lower row: sSFR. In each row, the three panels from left to right are: best--fit values of all spaxels with S/N$\geq$3, including bulge spaxels; arm spaxels; interarm spaxels, with the bulge excluded.}
    \label{fig:ngc628ParamMaps}
\end{figure}

\begin{figure}[!h]
    \centering
    \includegraphics[width=1\linewidth]{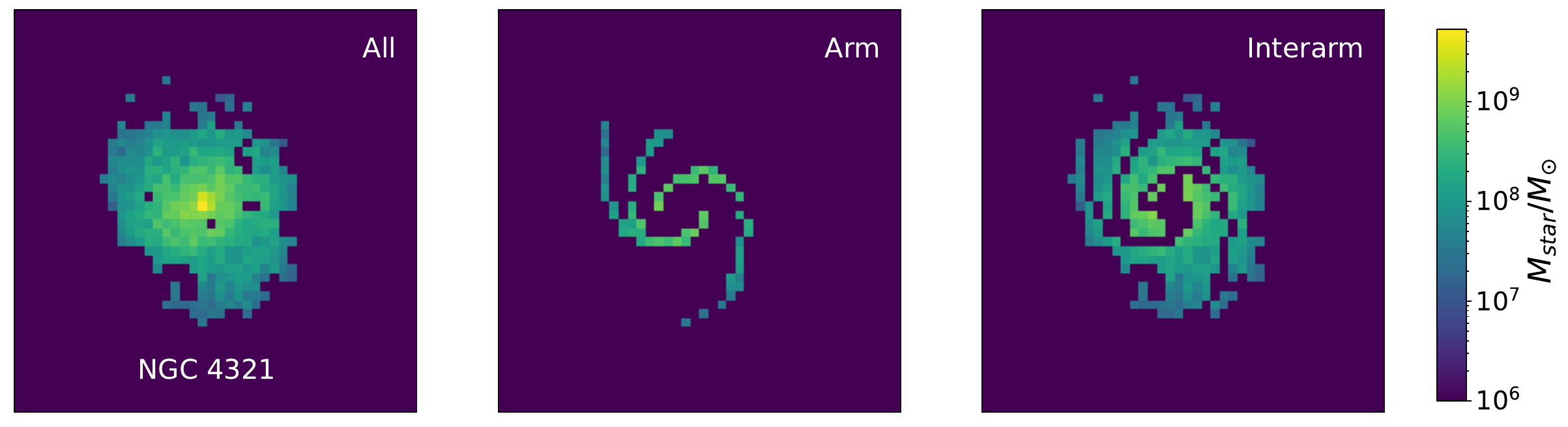} 
    \includegraphics[width=1\linewidth]{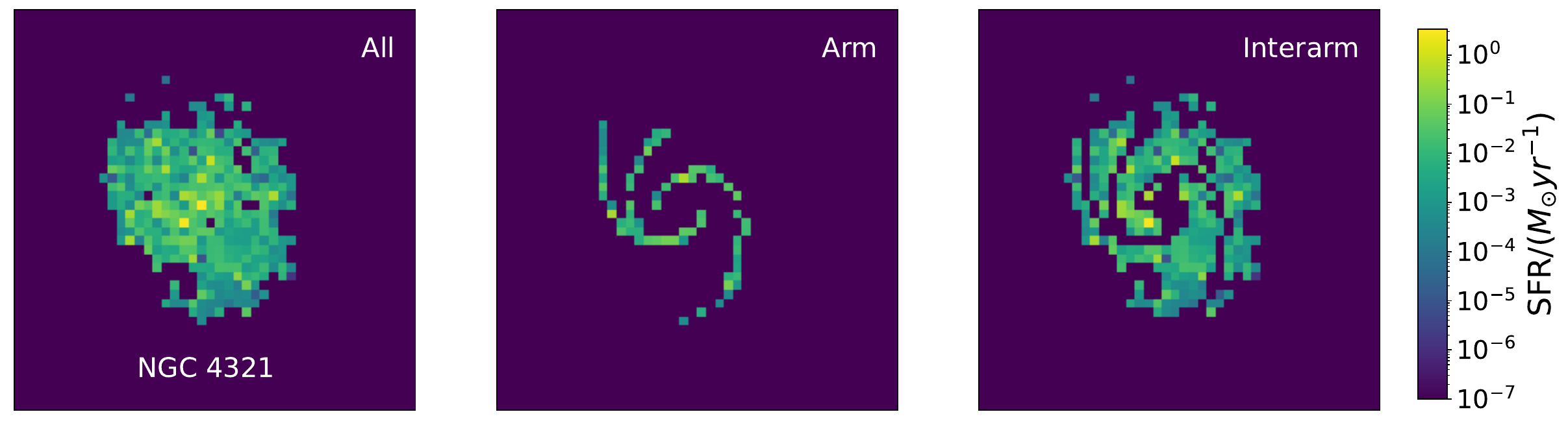} 
    \includegraphics[width=1\linewidth]{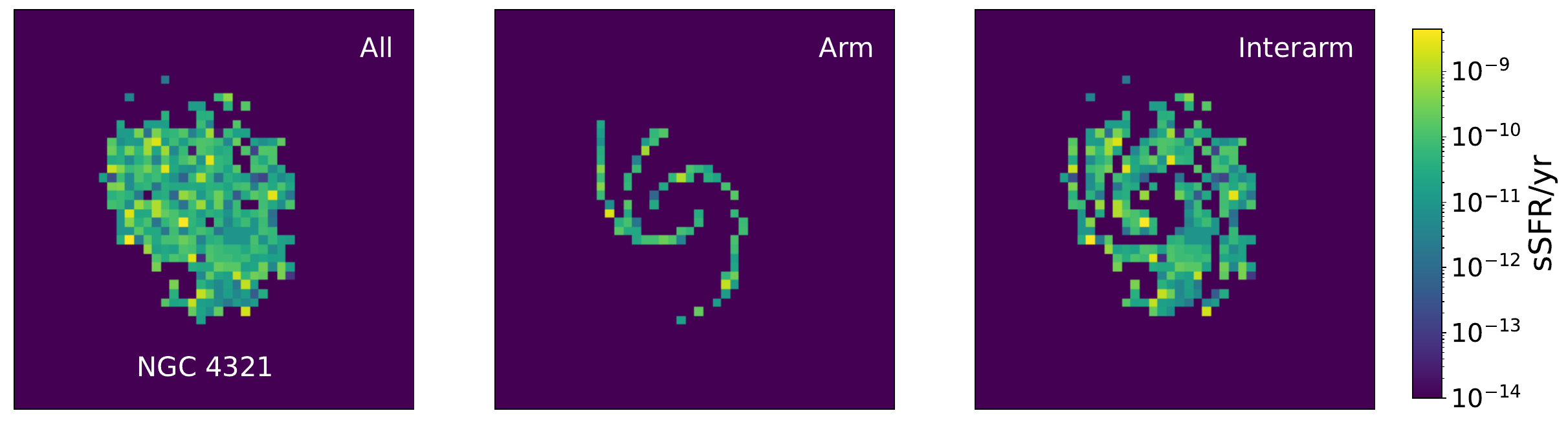} 
    
    \caption{Same parameter maps as in Figure~\ref{fig:ngc628ParamMaps}, but for NGC 4321.}
    \label{fig:ngc4321ParamMaps}
\end{figure}

After fitting the SEDs of all spaxels in each galaxy, we generate maps of the best-fit physical parameters. Figures  \ref{fig:ngc628ParamMaps} and \ref{fig:ngc4321ParamMaps} show how the best-fit stellar mass M$_\textup{star}$, SFRs, and sSFRs distribute across the whole extent of NGC 628 and NGC 4321, respectively. The spaxel values reported in these maps, as well as gas masses calculated from CO maps (see below),  are the basis for all results discussed in this section.

We note that we use the direct outputs from MAGPHYS, stellar mass and SFR, as opposed to their surface brightness equivalents (parameter/area). The reason is because the key parameters we are interested in, sSFR, SFE and t$_{form}$, are insensitive to this choice. For completeness, the reader interested in the surface brightness equivalents of M$_{star}$, M$_{gas}$ and SFR per unit  of kpc$^2$, will want to add 0.14~dex and subtract 0.30~dex to the logarithmic values of the parameters in NGC\,628 and NGC\,4321, respectively.

We first perform our analysis on the global values for each galaxy, and then move to a radial analysis, to identify how galactocentric gradients can affect our results.

\subsection{Specific SFR and $t_{form}$}
We quantify how the distributions of sSFR and  the ages of the luminosity--weighted oldest stellar populations 
distribute across each galaxy, separating the  arm and interarm regions.  We choose these two parameters because the sSFR has been used in previous works as a discriminant for the `gatherer' versus `trigger' arm scenarios, while the ages of the luminosity--weighted oldest stellar populations provide us with a simple metric to gauge potential differences in bulk ages between arm and interarm regions. In other words, if spiral arms are triggers of star formation, their luminosities will be dominated by the emission from the youngest stellar populations; since SED ages are luminosity--weighted, this will skew the mean ages of the spiral arms to younger values than those of the interarm regions. One caveat to the oldest population age parameter is that it is generally challenging to constrain from SED fitting of broad--band photometry, and can depend on the adopted SFH. Thus we will use this parameter only in conjunction with other parameters, as a sanity check.

Here and in all subsequent histograms, we plot the distribution of best--fit parameters from the SED fits of the spaxels in each galaxy, separately for arm and interarm regions, as follows. The probability density ($y$-axis) in the $i$-th bin is defined as $N_i/(N_{\rm tot}\Delta x)$, where $N_i$ is the number count in the bin, $N_{\rm tot}$ is the total number count, and $\Delta x$ is the bin width.

\begin{figure}[!h]
    \centering
    \includegraphics[width=8.5cm]{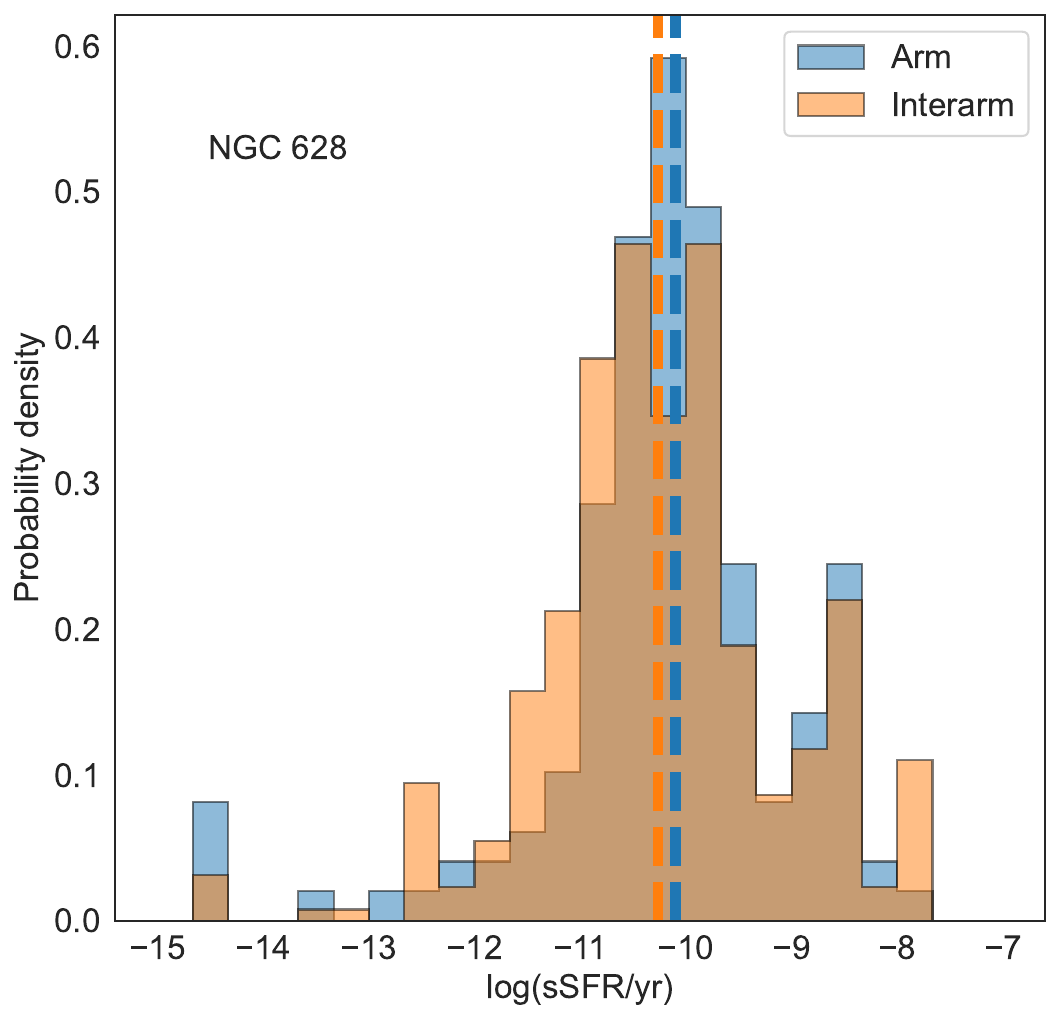}
    \includegraphics[width=8.5cm]{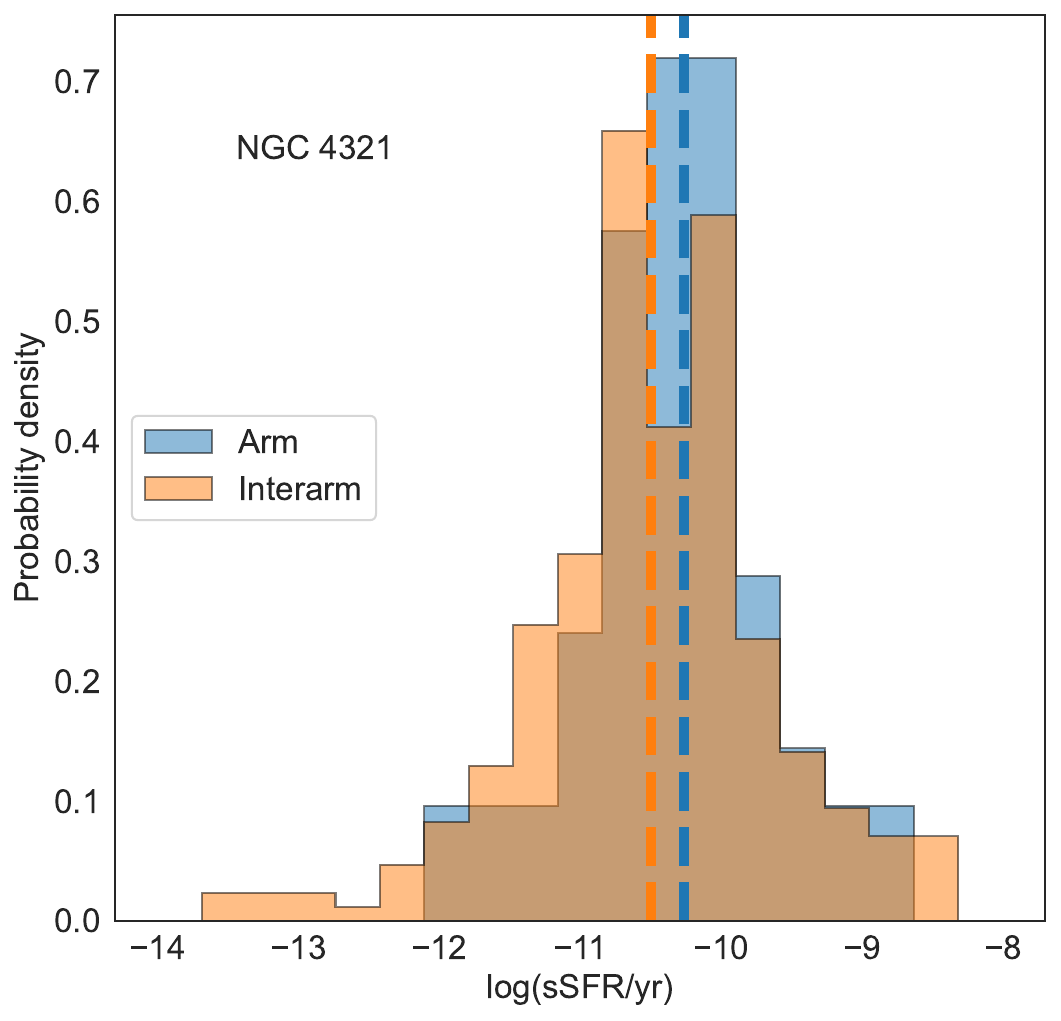}
    \caption{Best--fit values distributions of the sSFR (where the SFR is calculated over the most recent $10^7$ yr) for NGC 628 (left) and NGC 4321 (right). The arm spaxels and the interarm spaxels are plotted with blue and orange color, respectively. 
    For both galaxies, the sSFR distributions of arm/interarm regions are similar and their median values are shown as vertical dash lines and listed in Table~\ref{table:ksTest},  which also lists the results of the two--sample KS tests comparing the arm/interarm distributions.}

\label{fig:sSFRhist}
\end{figure}

The sSFR distributions of the best--fit values for the arm/interarm spaxels are shown in Figure~\ref{fig:sSFRhist}, panel (a) and (b) for NGC 628 and NGC 4321, respectively. The median of the arm regions distribution is about 1.5--1.7 times higher than the median of the interarm region distribution, for both galaxies. Results are summarized in Table \ref{table:ksTest}, together with all medians and ratios of the medians for all parameters discussed in the section.

\begin{figure}[!h]
    \centering
    \includegraphics[width=8.5cm]{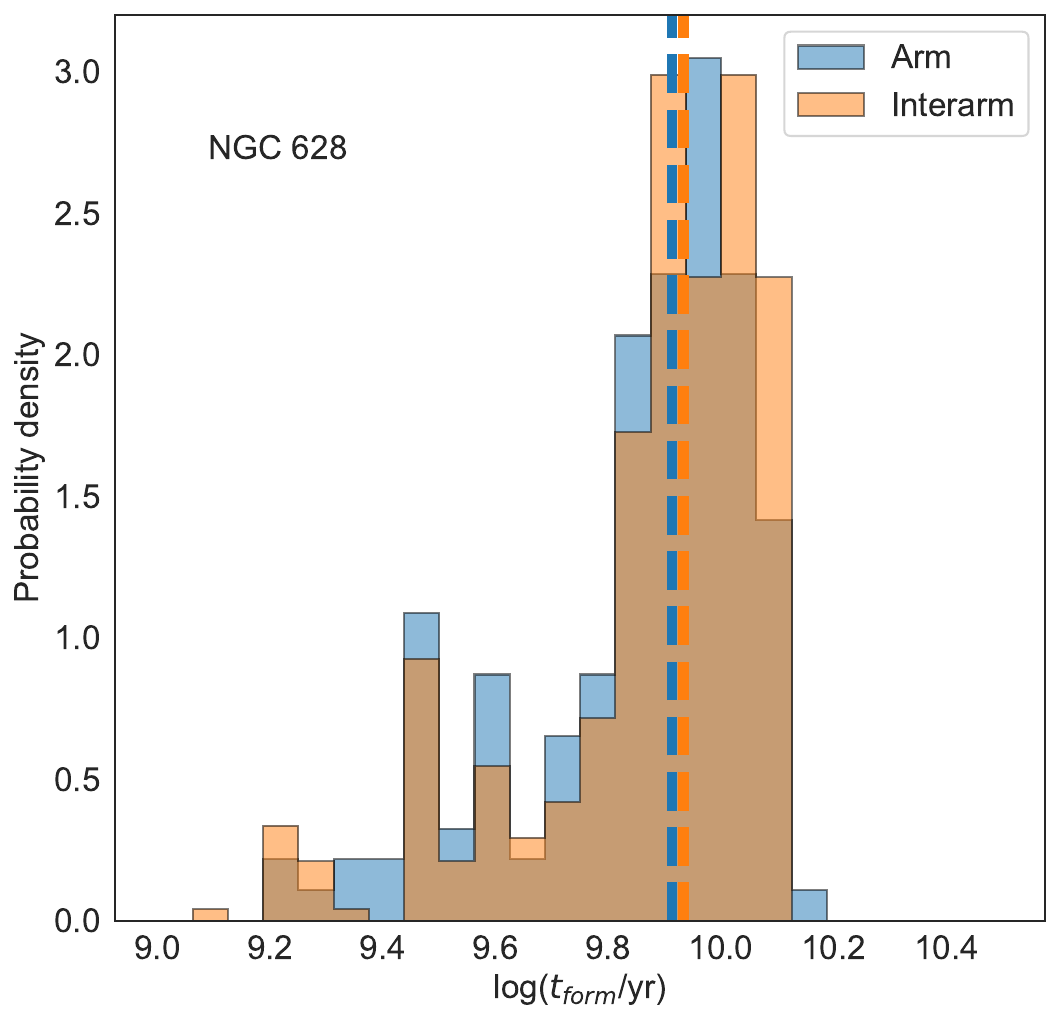} 
    \includegraphics[width=8.5cm]{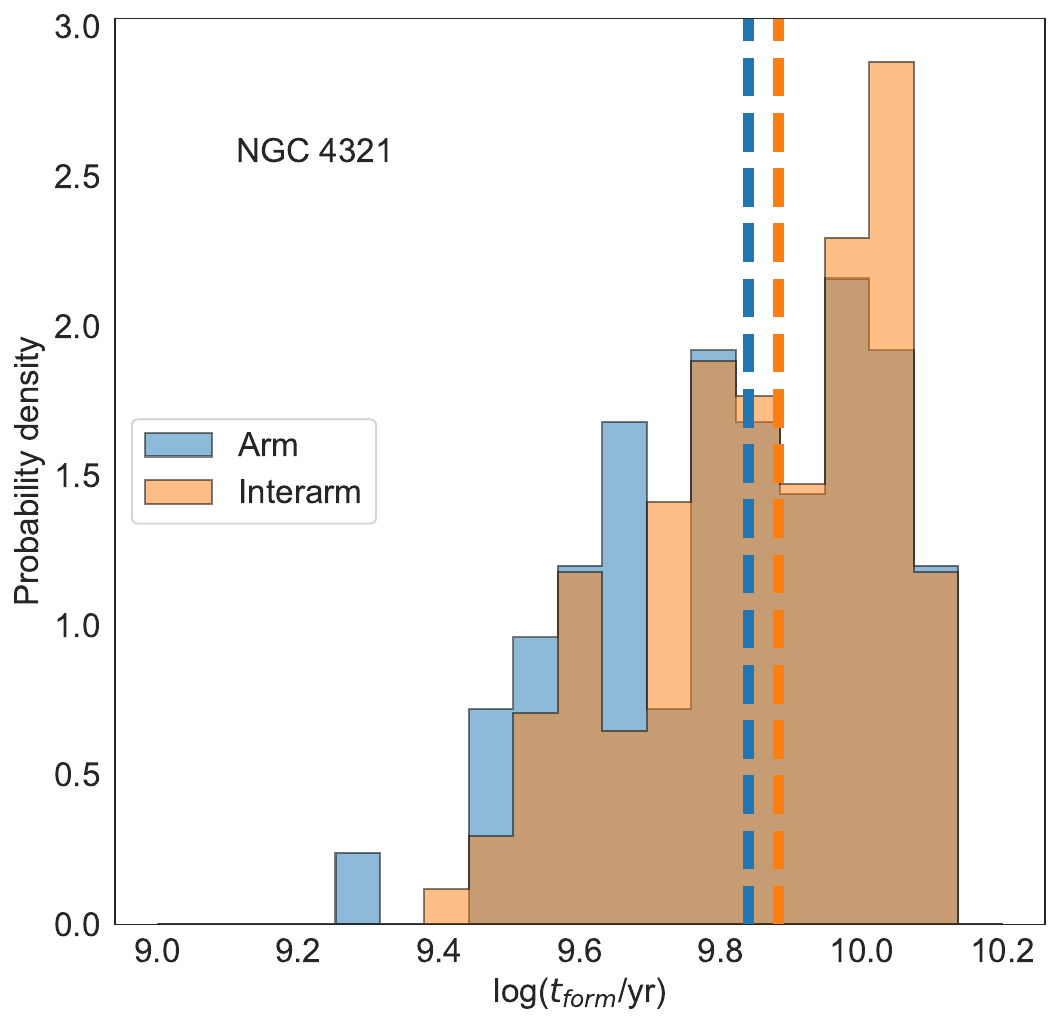} 
    \caption{Similar to Figure~\ref{fig:sSFRhist}, but for
    $t_{\text{form}}$, i.e. for the age of the luminosity--weighted oldest stellar population in the galaxy/regions being fitted. Left panel: NGC 628. Right panel: NGC 4321. For both galaxies, the $t_{\text{form}}$ distributions of arm/interarm regions are similar  (Table~\ref{table:ksTest}). }
    
    \label{fig:tformHistogram}
\end{figure}

Figure \ref{fig:tformHistogram} shows the distributions of $t_\textup{form}$ - the age of the luminosity--weighted oldest stellar population from the best--fits of arm and interarm regions in each galaxy. The medians of the distributions of the spaxels in the arms are virtually indistinguishable from those of the interarms, for both galaxies. 

To better quantify the differences in the distributions of parameter values between arm and interarm regions, we perform two sample Kolmogorov–-Smirnov test (two-sample K–S test)\footnote{We use the Python package scipy.stats.ks\_2samp() to perform this test.}. 
From the two p-values of the sSFR distributions, 0.064 and 0.050, respectively, we conclude that the two distributions are within $\approx$2~$\sigma$ of each other for both galaxies, and we cannot rule out the null hypothesis (that the two sSFRs are drawn from the same distribution)  to this level of confidence.

The results are even clearer for the distributions of the oldest ages. From their two p-values we can not reject the null hypothesis to about 1~$\sigma$, implying that   the distributions of $t_\textup{form}$ for the arm and interarm regions are very similar to  each other. Thus, while the null hypothesis may still be marginally acceptable for sSFR, it is more strongly supported when combined with the results for $t_\textup{form}$. 

\begin{figure}
    \centering
    \includegraphics[width=8.5cm]{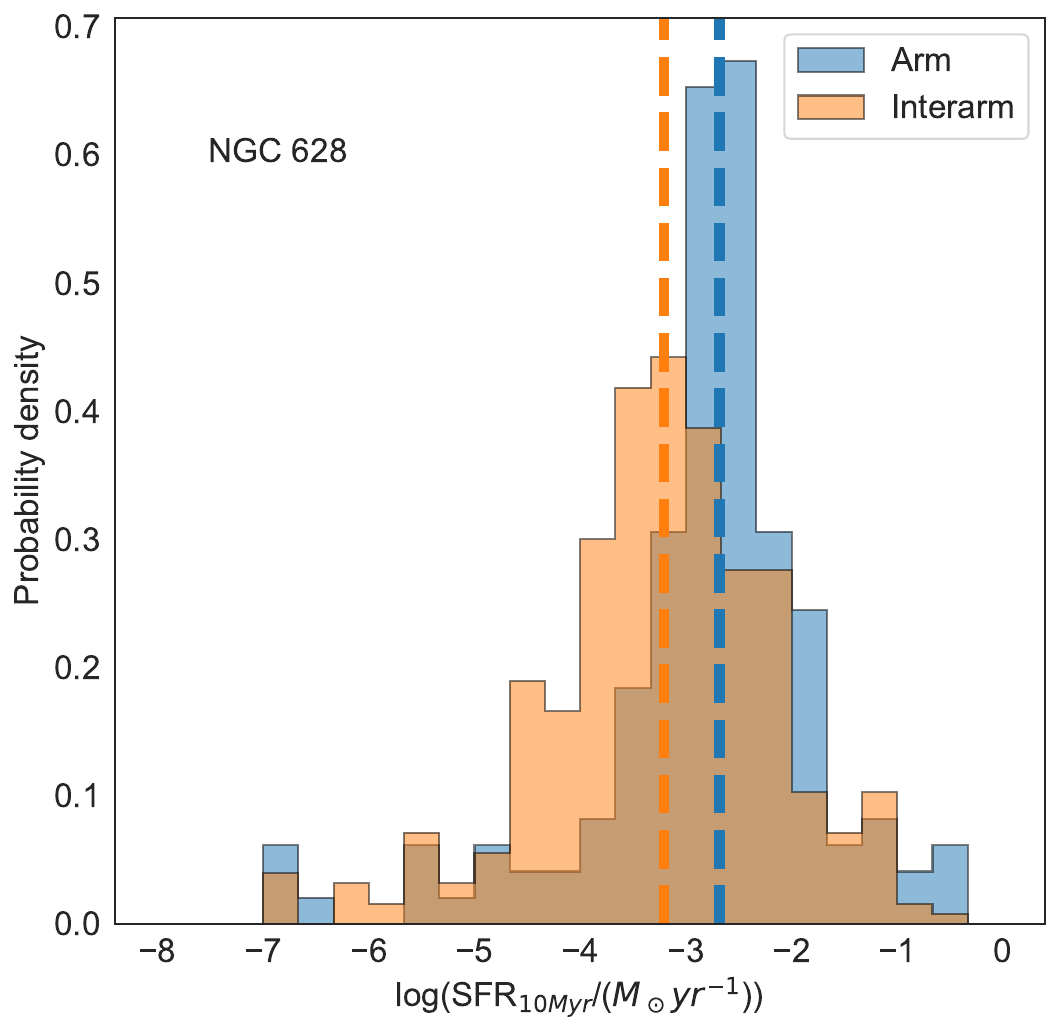} 
    \includegraphics[width=8.5cm]{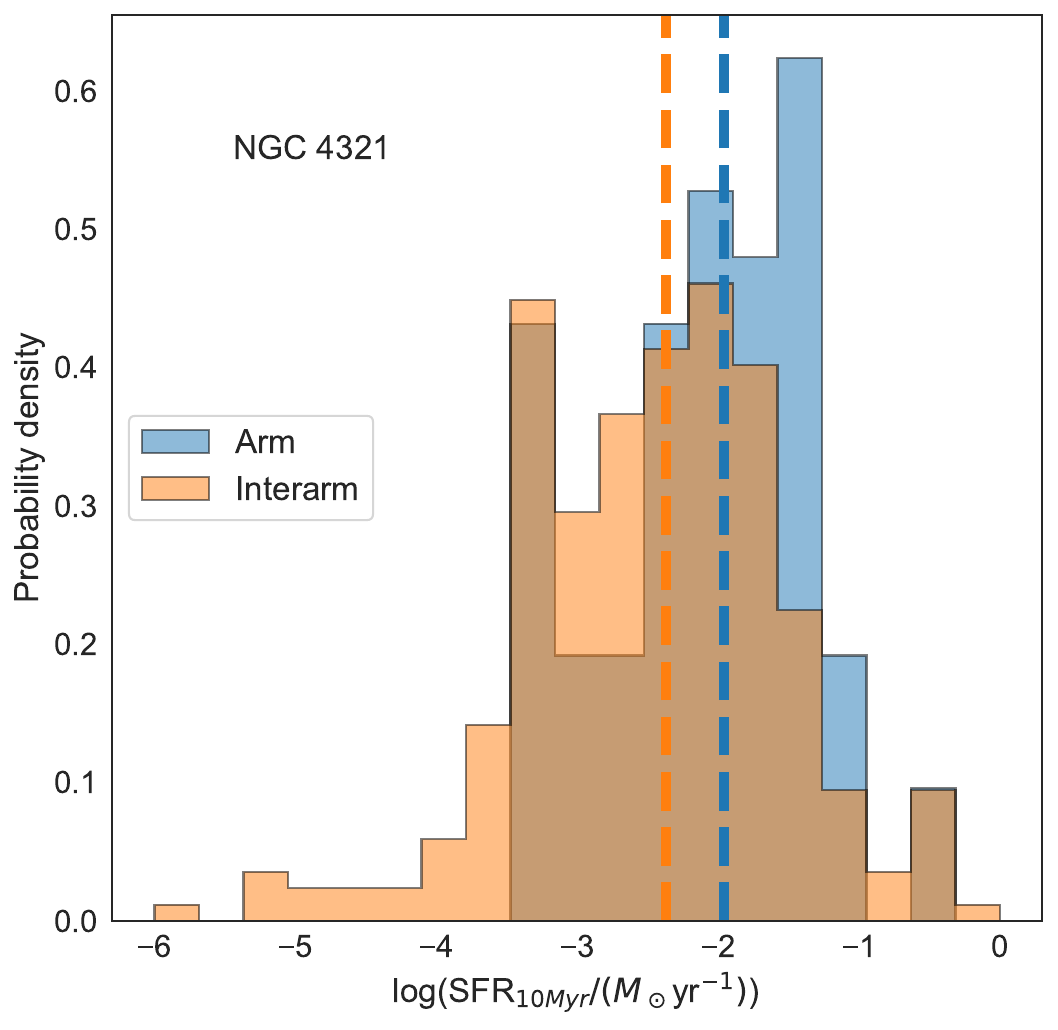}
    \includegraphics[width=8.5cm]{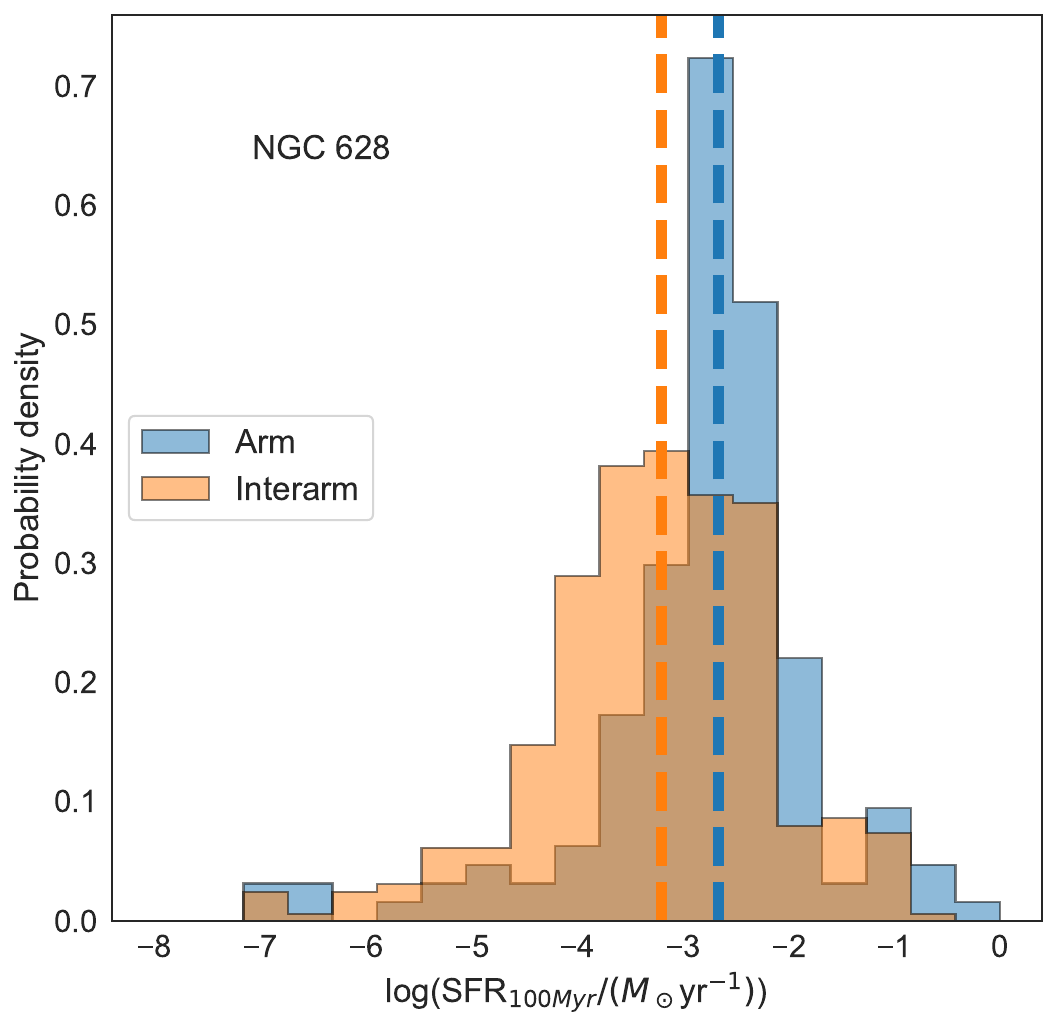} 
    \includegraphics[width=8.5cm]{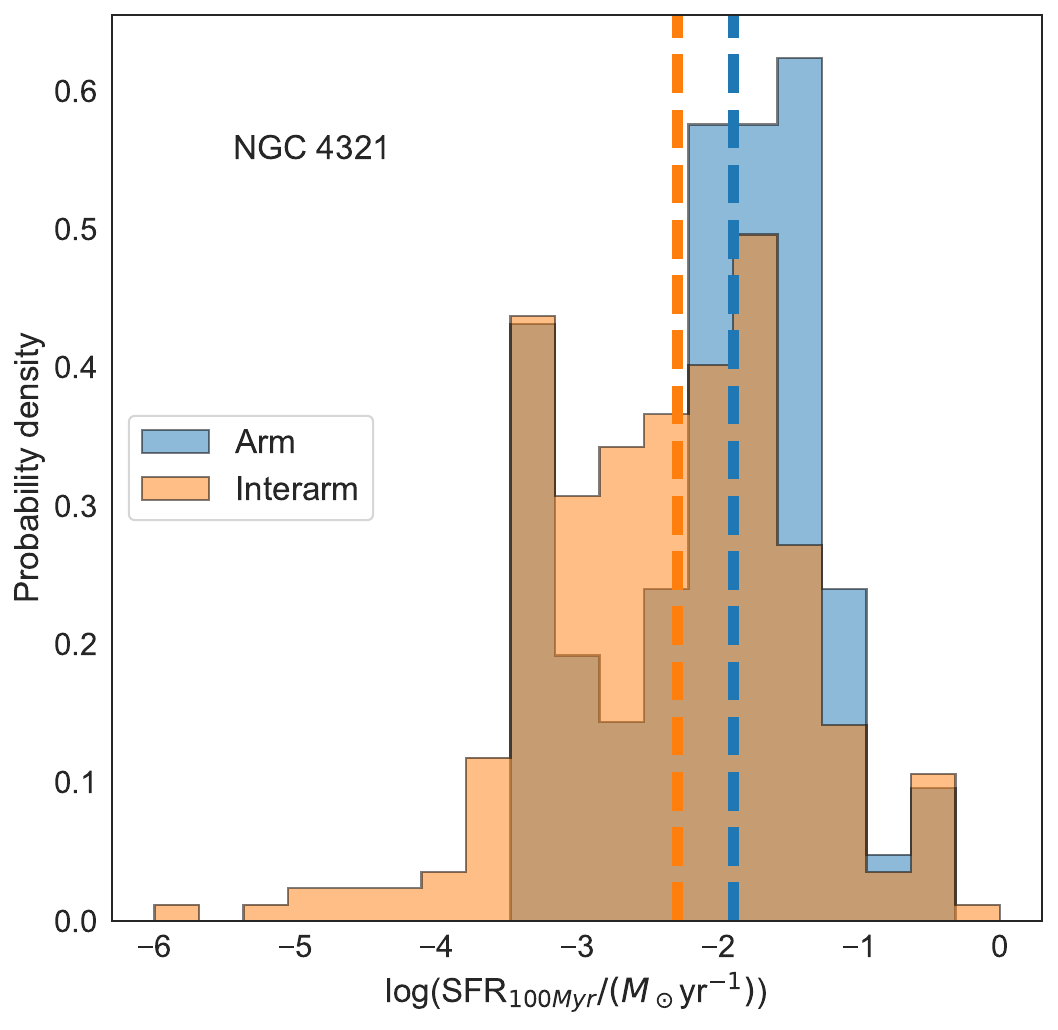}

    \caption{Similar to Fig.\ref{fig:sSFRhist} but for the SFR distributions. Upper row: the SFR averaged over the most recent $10^7$ yr. The median values are listed in Table~\ref{table:ksTest}. Lower  row: the SFR averaged over the most recent $10^8$ yr, whose distributions in log(SFR$/M_\odot yr^{-1})$ have median values   
    $-2.665$ ($-3.204$), and $-1.891$ ($-2.288$) for the arm (interarm) spaxels, for NGC\,628 and NGC\,4321, respectively. The difference in the medians between SFR$_{10Myr}$ and SFR$_{100Myr}$ is within 20\% for both galaxies.}
    \label{fig:SFRhist}
\end{figure}

\begin{figure}[!h]
    \centering
    \includegraphics[width=8.5cm]{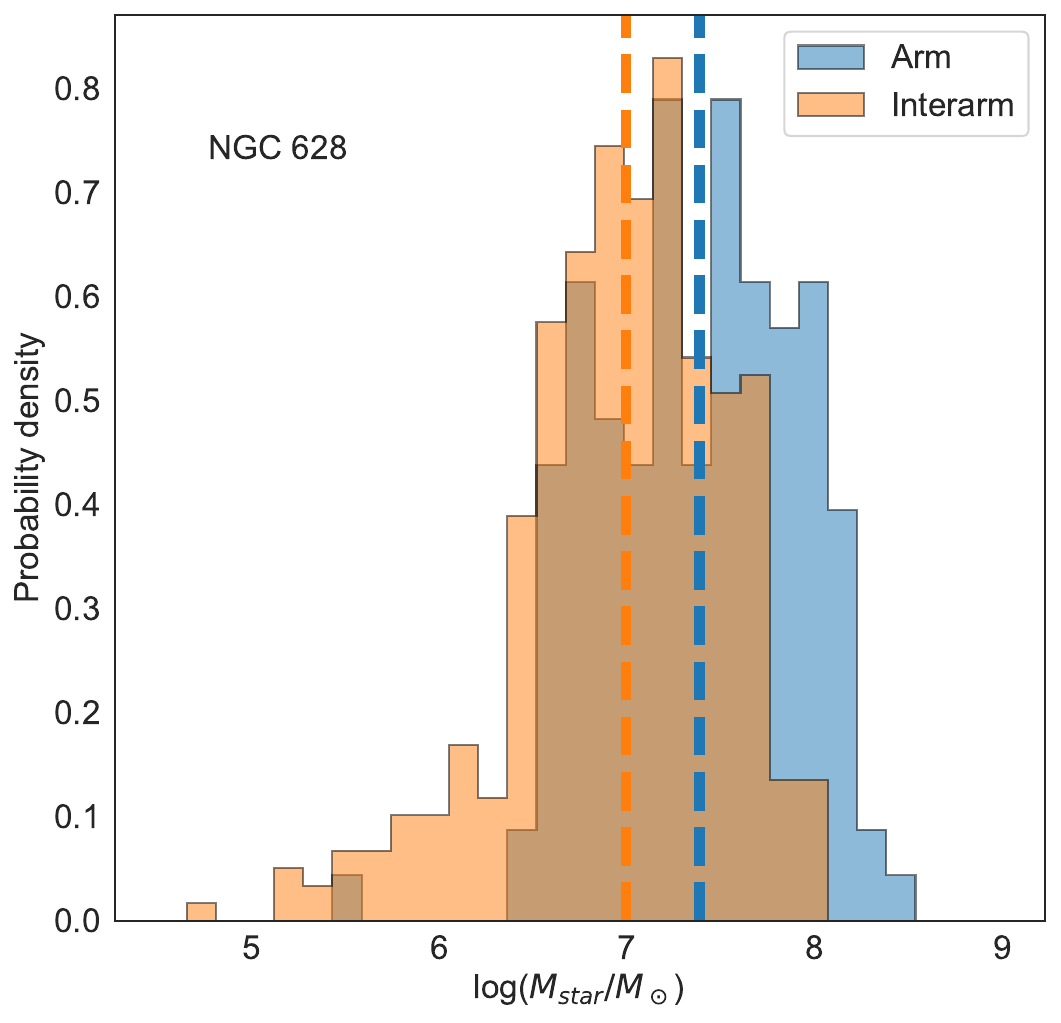}
    \includegraphics[width=8.5cm]{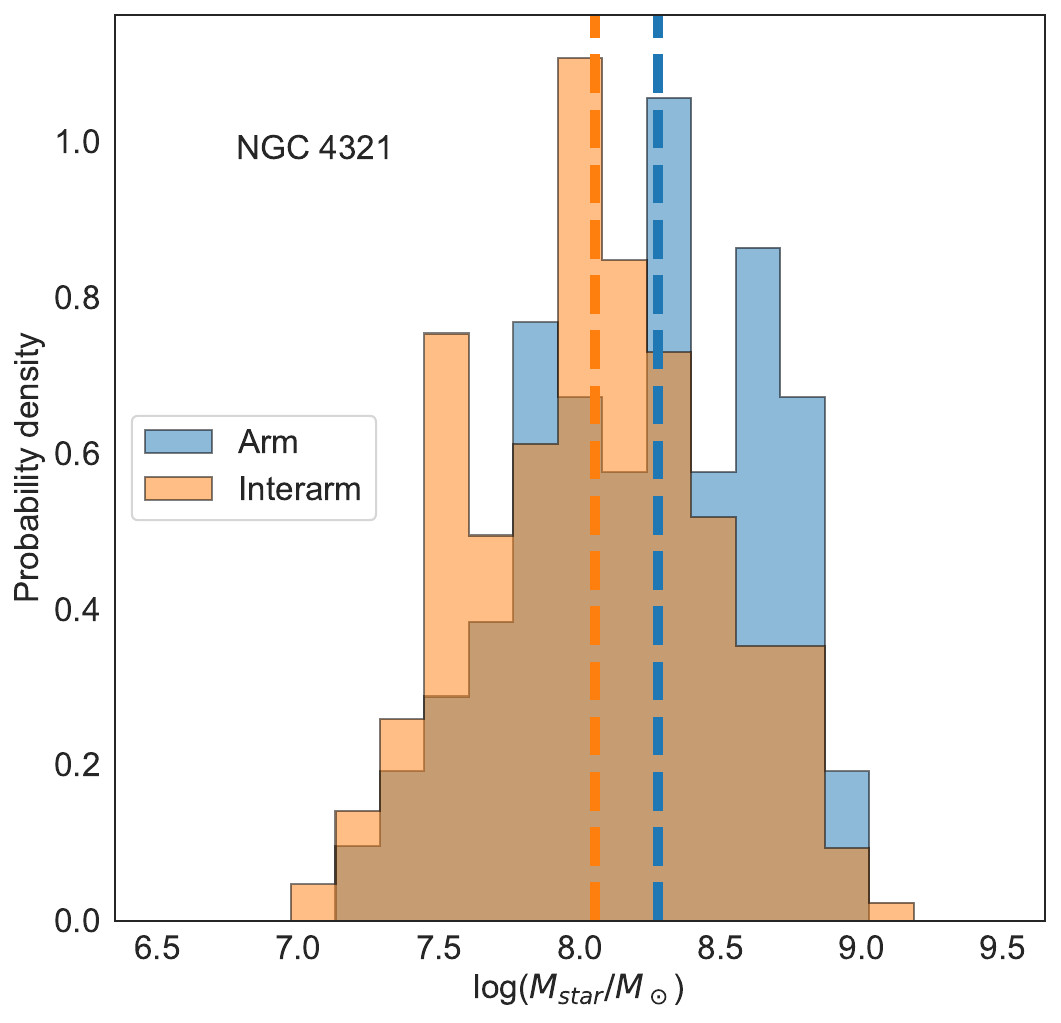} 
    \caption{ Similar to Fig.\ref{fig:sSFRhist} but for stellar mass. For both galaxies, the stellar masses are larger in the arm regions on average (Table~\ref{table:ksTest}).}   
    \label{fig:Mstarhist}
\end{figure}

We show in Figures~\ref{fig:SFRhist} the distributions of the SFRs averaged over the most recent $10^7$ years (top row) and over the most recent $10^8$ years (bottom row), 
separately for arm and interarm regions for the two galaxies. We readily observe that the arm regions have, on average, 2.6--3.4 times higher SFRs than the interarm regions in both galaxies, as also summarized in Table~\ref{table:ksTest}.
In addition, we observe that, 
in each galaxy and region (arm or interarm), differences between SFR$_\textup{10Myr}$ and SFR$_\textup{100Myr}$ are within 20\%, lending robustness to our results.  Stellar masses are also larger, on average, in the arm regions than the interarm regions by factors 1.7--2.5, as shown in Figure \ref{fig:Mstarhist} (see, also, Table~\ref{table:ksTest}).

In summary, while differences in the medians of the SFR and stellar mass distributions between arm and interarm regions are  significant, the differences in the medians of the sSFR are less than a factor of 2 in both galaxies and do not appear to be statistically significant.

\subsection{Star Formation Efficiency}
\begin{figure}[!h]
    \centering
    \includegraphics[width=8.5cm]{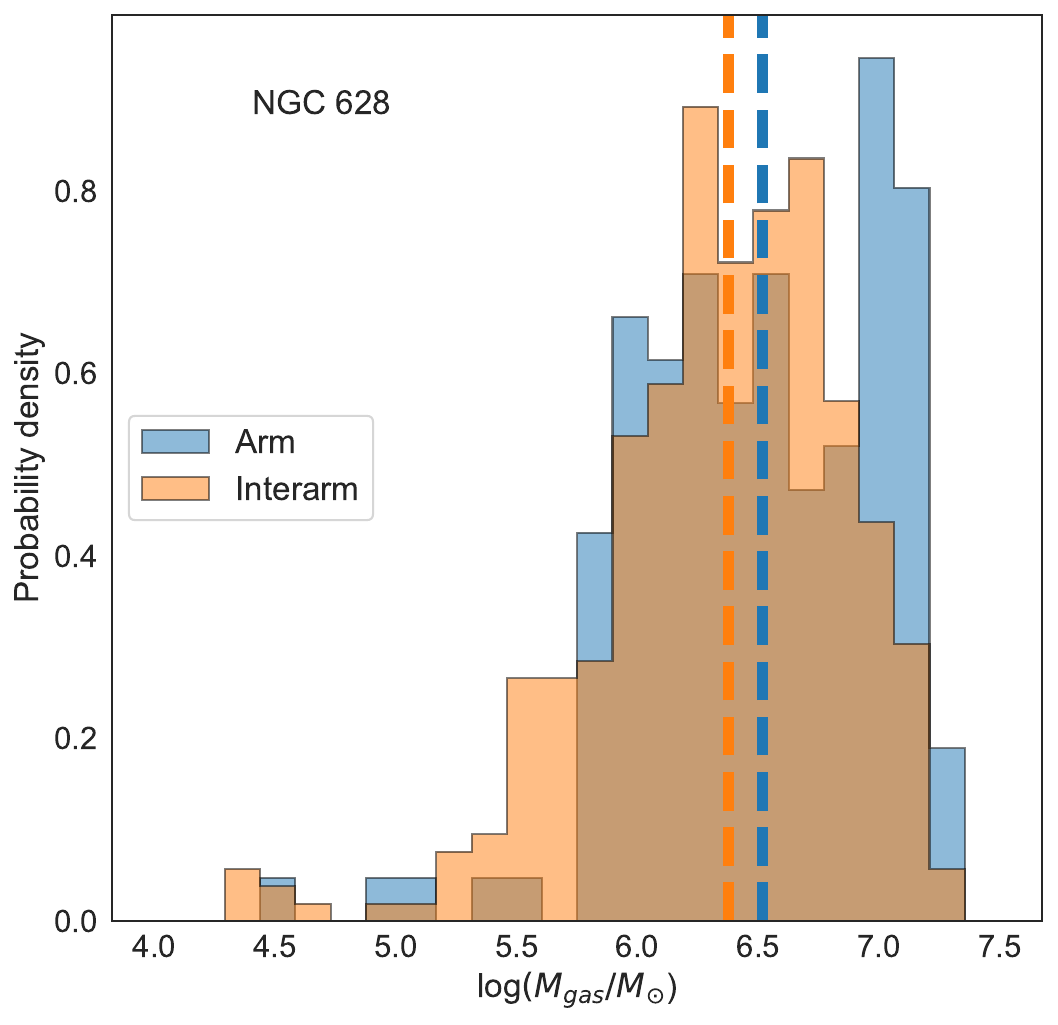}   
    \includegraphics[width=8.5cm]{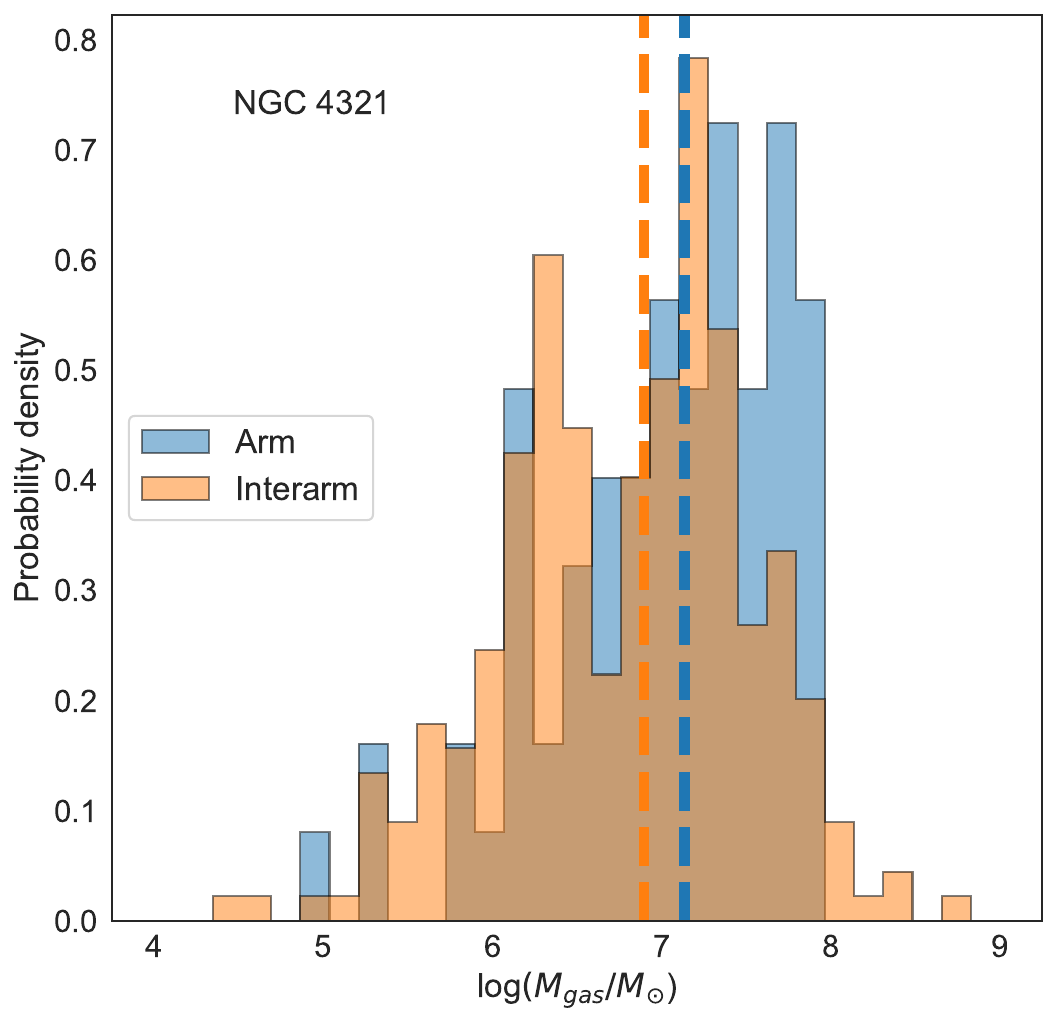}
    
    \caption{Similar to Fig.\ref{fig:sSFRhist} but for the molecular gas mass distributions of NGC 628 (left) and NGC 4321 (right).  
    In both galaxies, the average gas masses are higher in the arm regions, indicating an enhanced concentration of molecular gas but less than a factor of 2.}
\label{fig:gasMassHist}
\end{figure}

Following the work of \cite{Foyle2010}, we use the CO $J=2\rightarrow 1$ maps from HERACLES \citep{Leroy+2009}  to estimate molecular gas masses ($M_\textup{gas}$) in each spaxel. For this analysis, we ony retain spaxels with uncertainties in the CO flux of $\approx$20\% or less. 
The conversion factors from CO $J=2\rightarrow 1$ to CO $J=1\rightarrow 0$, $R_{21}$, used here is 0.65 \citep{Leroy+2013}. 
We adopt the CO-to-H$_2$ conversion factor $\alpha_\textup{CO,MW} = 4.3 M_\odot$ (K km s$^{-1}$ pc$^2$)$^{-1}$ \citep{Bolatto+2013}, a commonly used approximation for the Milky Way and appropriate for our two metal--rich galaxies \citep[both have about solar abundance, ][]{Moustakas+2010}.
While the conversion factor may introduce some uncertainty, it should have minimal impact on the relative difference between the arm and interarm regions, which is the focus of our analysis. The derived gas mass distributions for the spaxels in the two galaxies, separated between arm and interarm regions, are shown in Figure \ref{fig:gasMassHist} with the medians and the results of the KS tests for the comparison between arms and interarms listed in Table~\ref{table:ksTest}. 

\begin{figure}[!h]
    \centering
    \includegraphics[width=8.5cm]{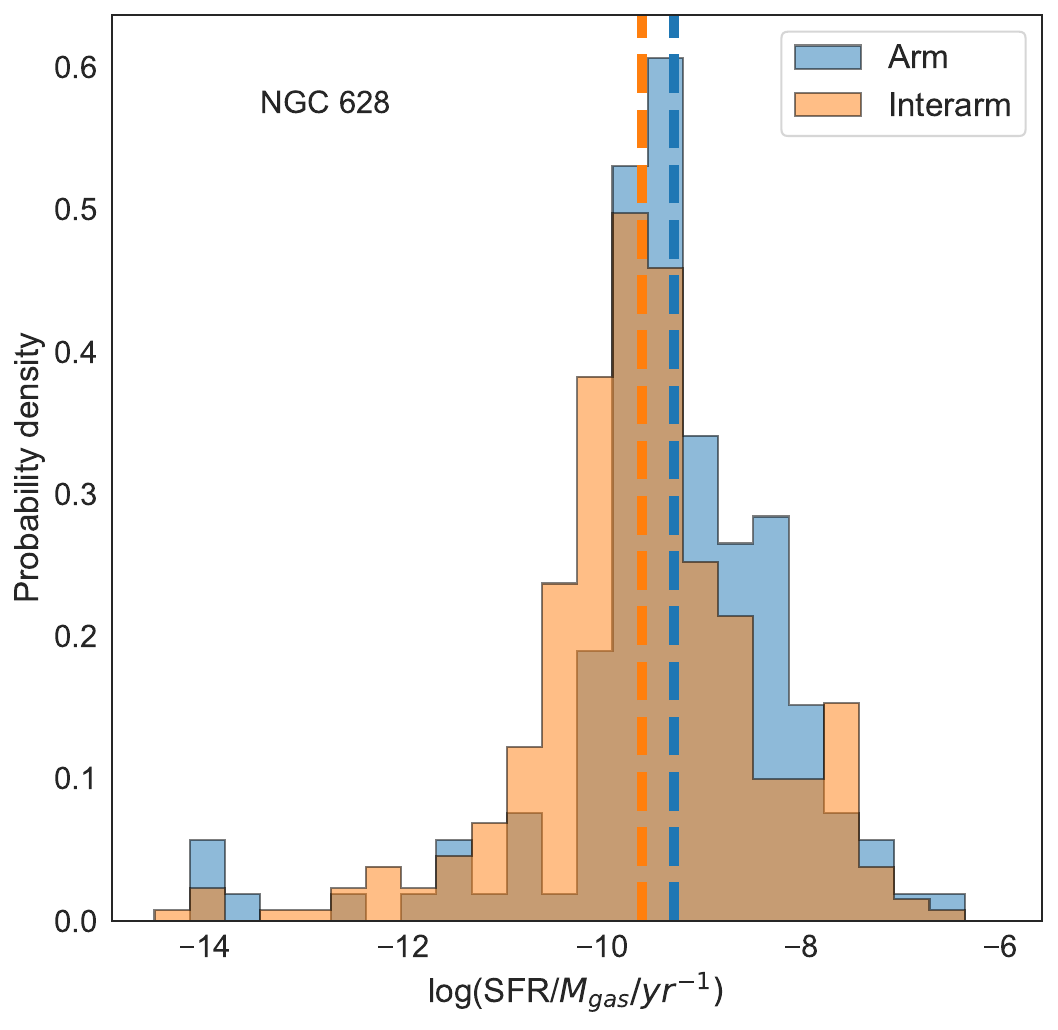} 
    \includegraphics[width=8.5cm]{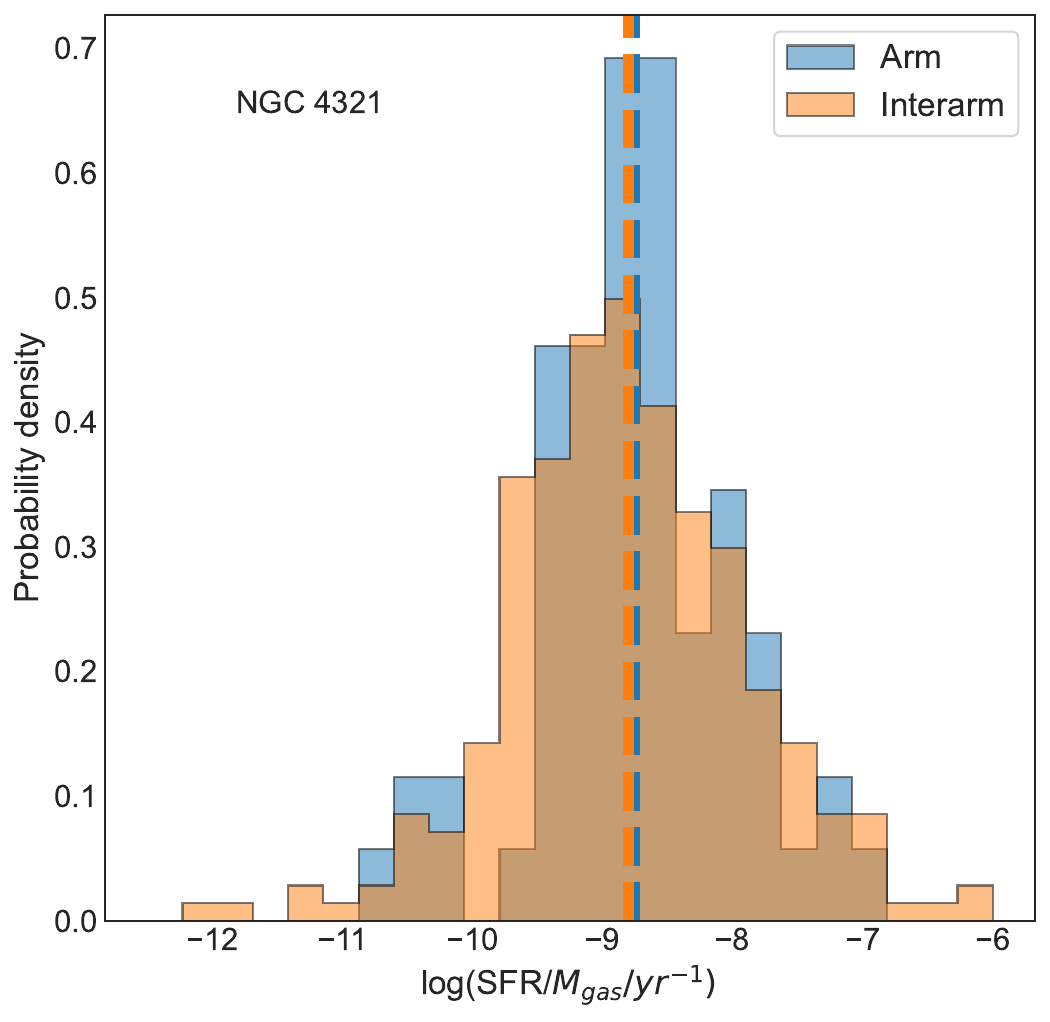}
    
    \caption{ The SFE, i.e., the ratio of the SFR, averaged over $10^7$ yr, to the gas mass. 
    Left: NGC~628. Right: NGC~4321. The differences in the arm/interarm medians are only significant for NGC~628 (Table~\ref{table:ksTest}). }
    \label{fig:SFR-mass-ratio-hist}
\end{figure}

Figure \ref{fig:SFR-mass-ratio-hist} shows the SFE, i.e., the ratio of the SFR to $M_\textup{gas}$. 
We find that in NGC 628, the median of the SFE distribution in the arm regions is 2.097 times the median in the interarm regions. The KS-test gives a p-value of $8.849\times 10^{-5}$, implying that the two distributions are different at the $\approx$3.85~$\sigma$ level (Table~\ref{table:ksTest}). Our result for NGC 628 is consistent with the results of \cite{Foyle2010} on the same galaxy, where they also observed a slightly larger SFE in the arm regions. Additionally, the shapes of the SFE distributions in our and their studies are similar to each other. 

For NGC 4321, the median SFE in arm regions is 1.1 times higher than the median in interarm regions.
And with a p-value of 0.368, the two distributions are not distinguishable.  

We conclude that the SFE enhancement in the arm regions is relatively small, about a factor 2 or less than the value of the arm regions. For NGC~628, the difference in the distributions of SFE between arm and interarm regions is significant, but it is not for NGC~4321.

\begin{table}[h!]
\centering
\begin{tabular}{ c|c|c|c|c|c|c|c } 
\hline
Galaxy & Parameter & Arm & Arm &  Interarm & Interarm & Median & p-value \\
 & & \#spaxels & median (log) & \#spaxels & median (log) & ratio & \\
\hline
NGC 628 & $M_\textup{star}$ & 147& 7.384 & 381 & 6.994 & 2.452  & $1.032\times 10^{-8}$\\
NGC 4321 & $M_\textup{star}$ & 66 & 8.276 & 269  & 8.053 & 1.671 &  $6.175\times 10^{-3}$\\
\hline
NGC 628 & SFR$_\textup{10Myr}$ & 147& -2.668 & 381 & -3.209 & 3.385 & $9.033\times 10^{-10}$ \\
NGC 4321 & SFR$_\textup{10Myr}$ & 66 & -1.957 & 269  & -2.370 & 2.587 & $1.588\times 10^{-3}$\\
\hline
NGC 628 & sSFR$_\textup{10Myr}$ & 147& -10.099 & 381 & -10.265 & 1.468 & 0.064 \\
NGC 4321 & sSFR$_\textup{10Myr}$ & 66 & -10.264 & 269  & -10.497 & 1.709 & 0.050\\
\hline
NGC 628 & $t_\textup{form}$ & 147& 9.913 & 381 & 9.933 & 0.955 &  0.163\\
NGC 4321 & $t_\textup{form}$ & 66 & 9.839 & 269  & 9.882 & 0.906 & 0.341\\
\hline
NGC 628 & M$_\textup{gas}$ & 145 & 6.516 & 361 & 6.375 & 1.383 & 0.001\\
NGC 4321 & M$_\textup{gas}$ & 64 & 7.135 & 259  & 6.894 & 1.740 & 0.0477\\
\hline
NGC 628 & SFE & 145 &  -9.319 & 361 &  -9.643 & 2.108 & $7.204 \times 10^{-5}$\\
NGC 4321 & SFE & 64 & -8.752 & 259 & -8.799 & 1.113 & 0.368\\
\hline
\end{tabular}
\caption{The medians, median ratios, and Two-sample KS test statistics of the best fit parameters for each galaxy, divided between arm and interarm spaxels. The third and fifth columns list the number of spaxels in each region. There are fewer spaxels available for M$_\textup{gas}$ and SFE, because the CO emission in some spaxels is not well detected.}
\label{table:ksTest}
\end{table}

\subsection{Radial Analysis}
It is well established that the surface density of stellar mass, SFR and other parameters depends on galactocentric radius \citep[e.g.,][]{Freeman1970, Kormendy+2004}. In addition, there is a higher proportion of interarm spaxels relative to arm spaxels in the outskirts than in the centers of galaxies, which can potentially bias the interarm medians to the lower surface density values of the outskirts; this can  enhance the contrast with the arm medians without representing a physical effect.

\begin{figure}
    \centering
    \includegraphics[width=0.45\linewidth]{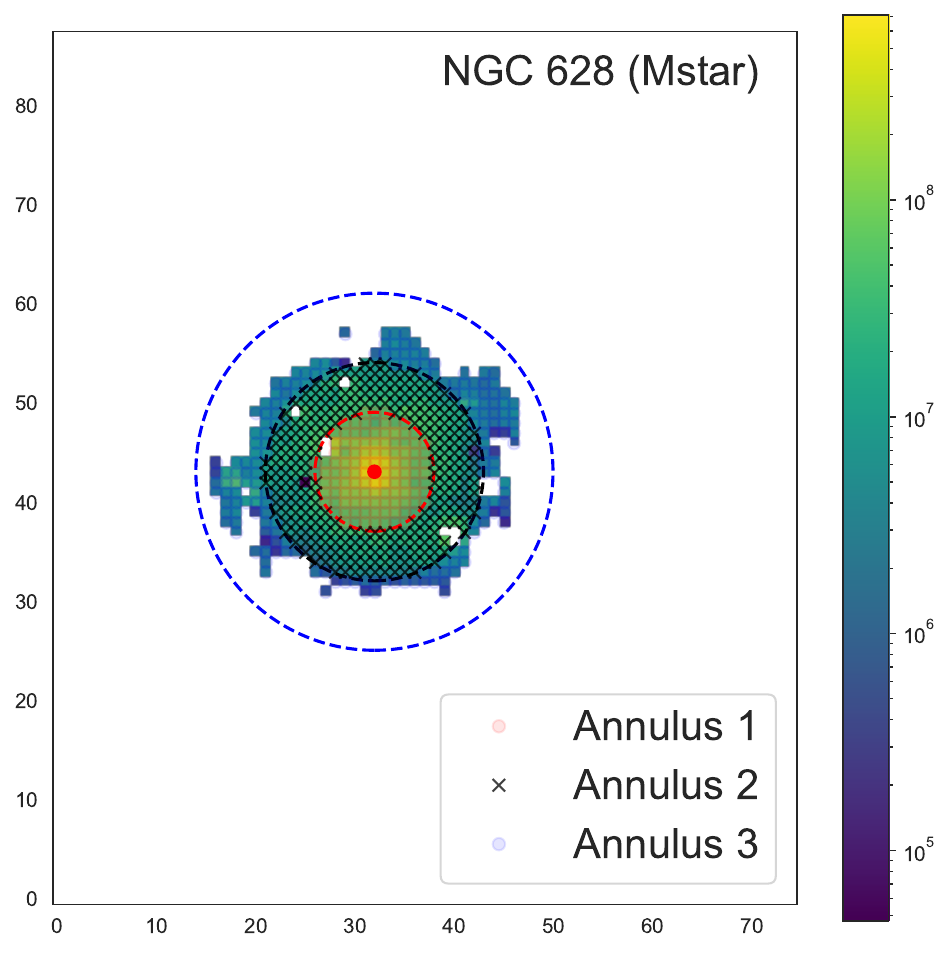}
    \includegraphics[width=0.47\linewidth]{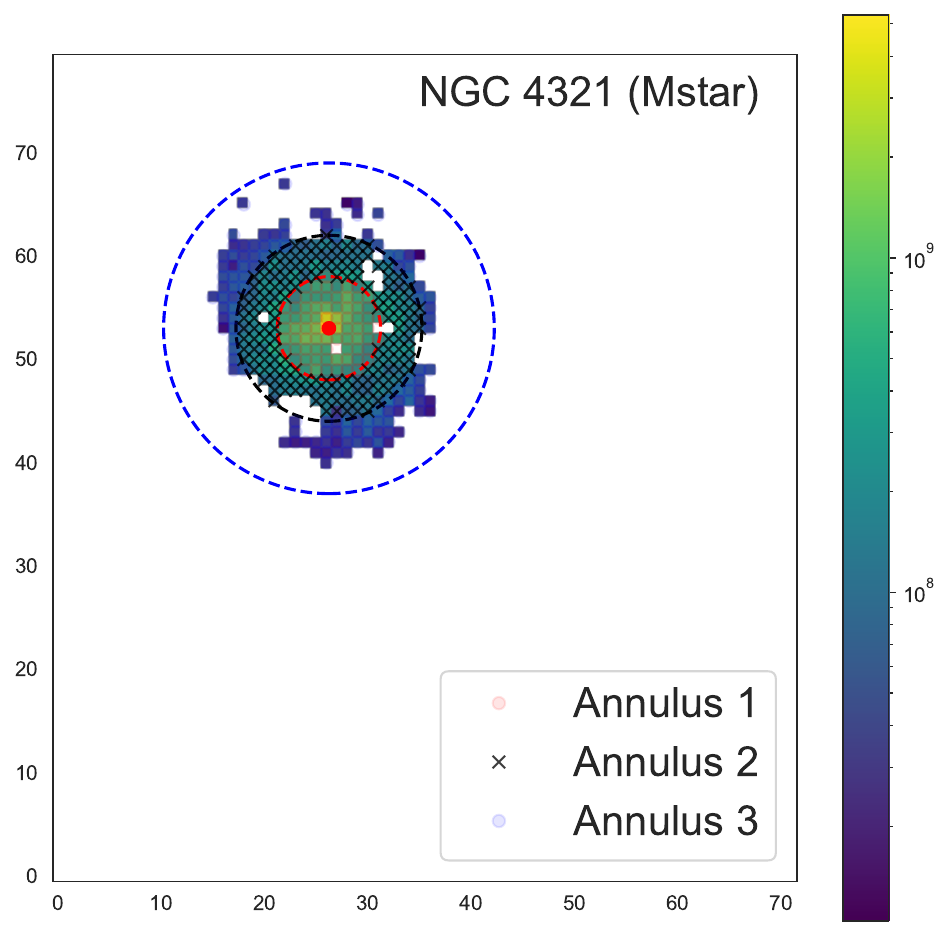}
    \caption{Three annuli plotted on the stellar mass maps of NGC 628 (Left) and NGC 4321 (Right). 
    For NGC~628, annulus R1 is 6 spaxels wide, R2 is 5 spaxels, and R3 is 7 spaxels. For NGC~4321, annulus R1 is 5 spaxels wide, R2 is 4 spaxels, and R3 is 7 spaxels. The numbers of arm and interarm spaxels in each annulus are summarized in Table \ref{tab:Mstar-Mgas-3annuli}.}
    \label{fig:ngc628-3annuli-location}
\end{figure}

\begin{figure}
    \centering
    \includegraphics[width=1\linewidth]{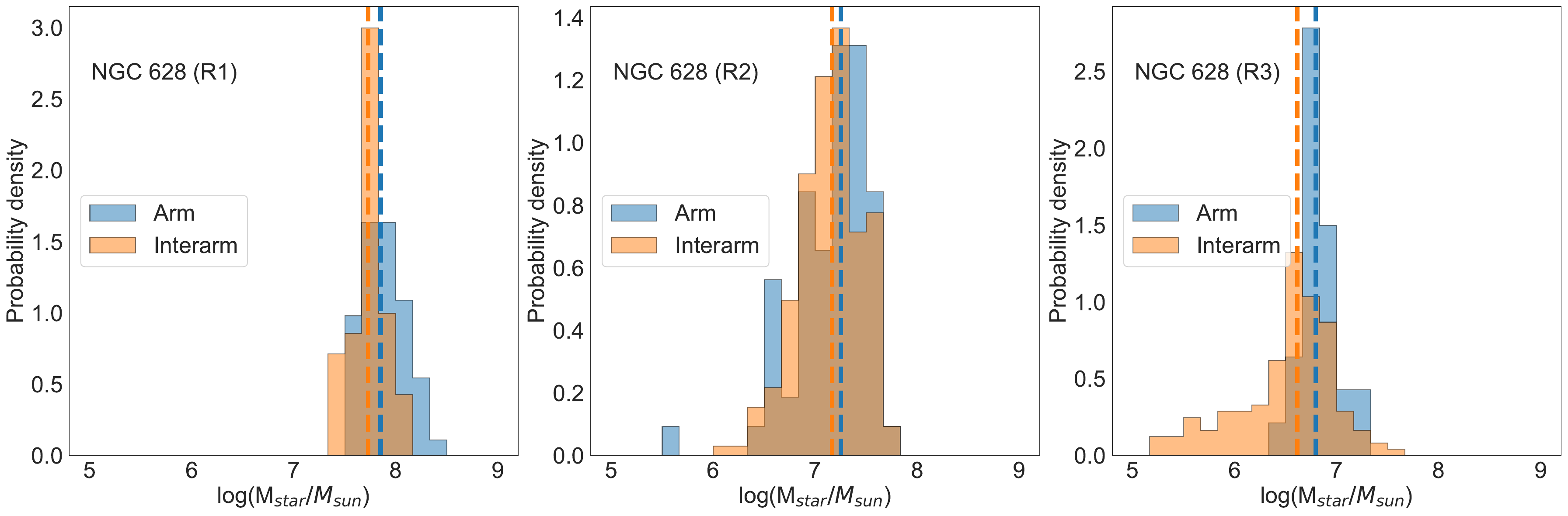}
    \includegraphics[width=1\linewidth]{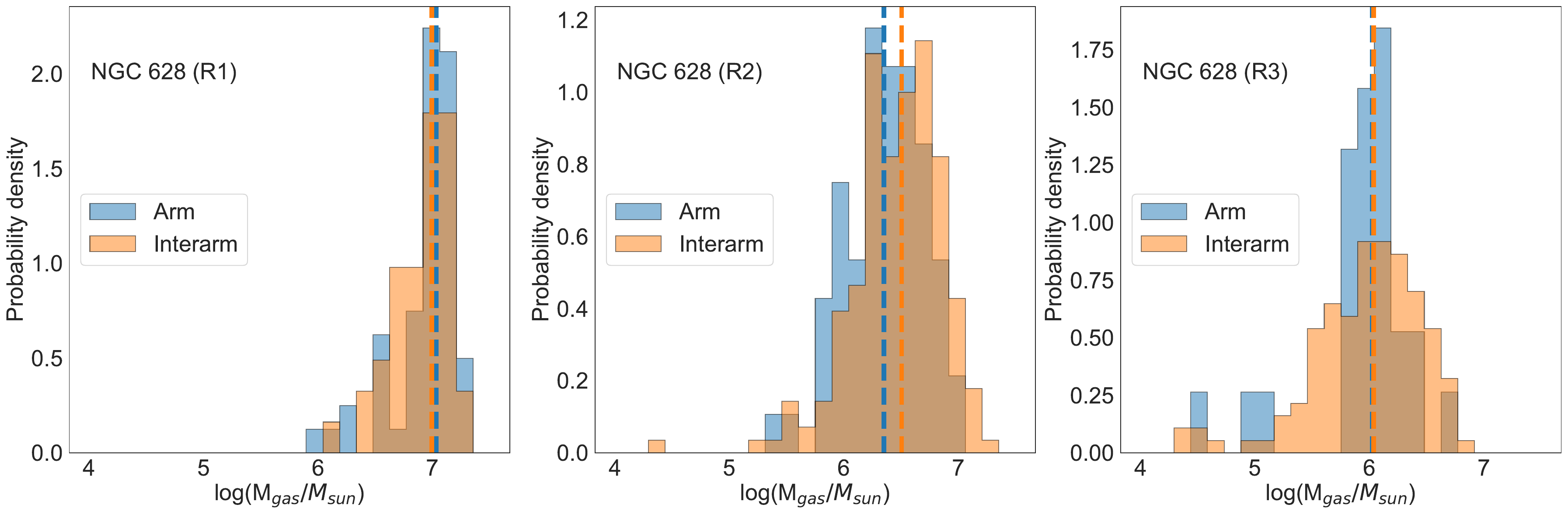}
    \includegraphics[width=1\linewidth]{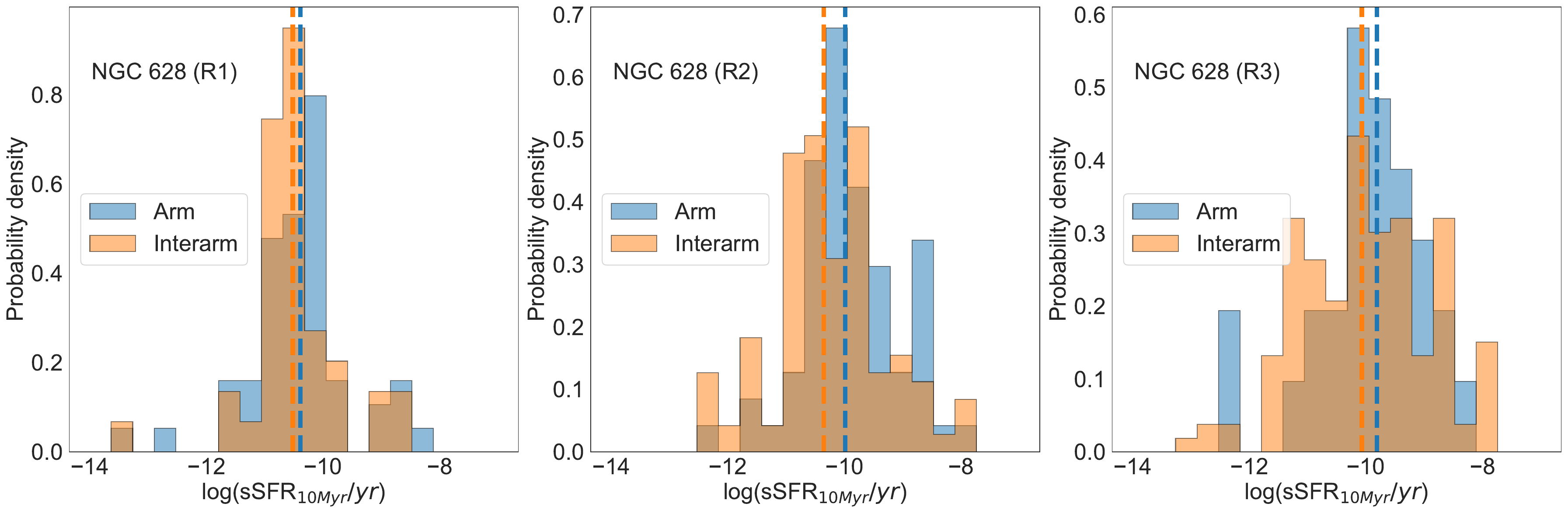}
    \caption{Log$(M_\textup{star}/M_\odot)$, log$(M_\textup{gas}/M_\odot)$ and sSFR distributions within the three annuli in NGC 628. In each annulus, the ratio of medians is smaller than the global median ratio.
    The numbers of arm and interarm spaxels and the values of the medians of the distribtions are listed in Table \ref{tab:Mstar-Mgas-3annuli}.}
    \label{fig:Mstar-hists-3annuli}
\end{figure}

\begin{figure}
    \centering
    \includegraphics[width=1\linewidth]{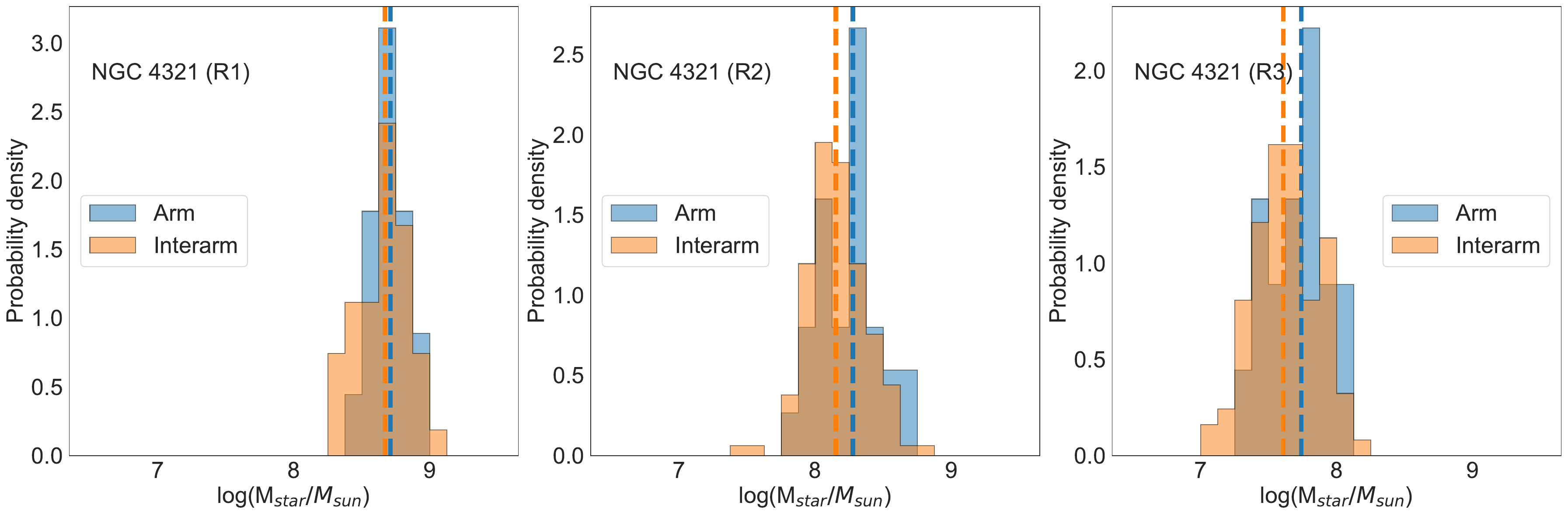}
    \includegraphics[width=1\linewidth]{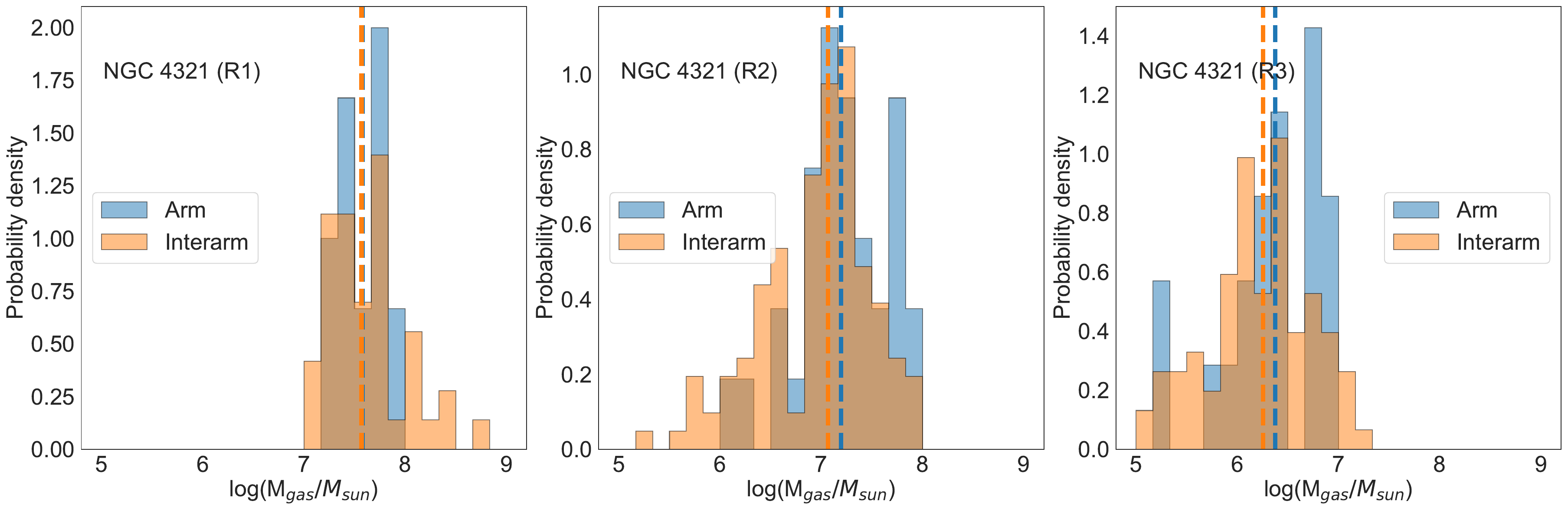}
    \includegraphics[width=1\linewidth]{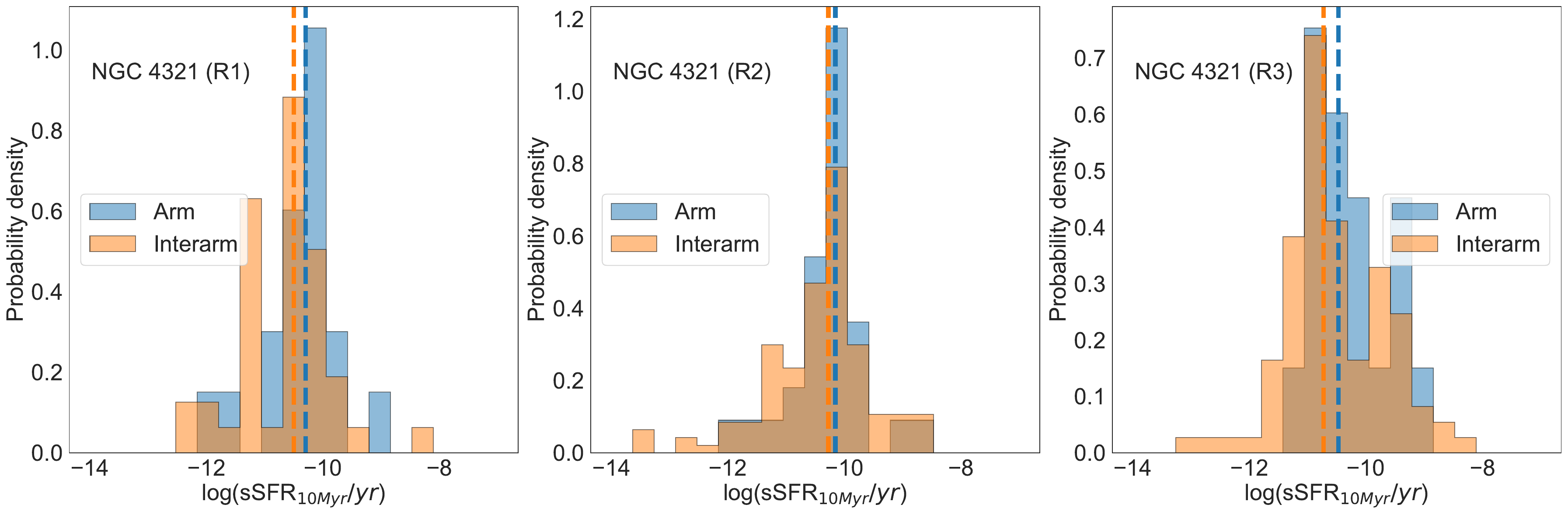}
    \caption{Same as Fig.\ref{fig:Mstar-hists-3annuli} but for NGC 4321.  
    The numbers of arm and interarm spaxels, the values of their medians and median ratios are listed in Table \ref{tab:Mstar-Mgas-3annuli}.}
    \label{fig:Mstar-hists-3annuli-ngc4321}
\end{figure}

To investigate the impact of galactocentric trends on our results, we divide each of our galaxies  into three annuli and repeat the analysis above as a function of galactocentric annulus. The number of annuli we divide each galaxy in is driven by the need to preserve statistical significance for the medians of spaxel distributions, which yields three annuli when  setting to 15 the minimum number of arm or interarm spaxels in an annulus. Figure \ref{fig:ngc628-3annuli-location} shows the annuli on the stellar mass maps of the two galaxies; the width of the annuli ranges from 4.2~kpc to 6~kpc in NGC\,628 and from 5.3~kpc to 9.3~kpc in NGC\,4321. Figures \ref{fig:Mstar-hists-3annuli} and \ref{fig:Mstar-hists-3annuli-ngc4321} show the  $M_{\textup{star}}$, $M_{\textup{gas}}$ and sSFR  distributions within each annulus across NGC 628 and NGC 4321, respectively. The median values and ratios are summarized in Table \ref{tab:Mstar-Mgas-3annuli} for the same list of parameters as Table~\ref{table:ksTest}.

The plots in Figures~\ref{fig:Mstar-hists-3annuli} and \ref{fig:Mstar-hists-3annuli-ngc4321} and the ratios of medians listed in Table~\ref{tab:Mstar-Mgas-3annuli} paint a slightly different picture for our physical parameters than what we have inferred from the global values. For both galaxies, the $M_{\textup{star}}$ contrast  between arm and interarm regions, as measured by the ratio of the medians of their spaxels distributions, decreases in the annuli  relative to the global values, suggesting that the radial trends matter for determining accurate contrasts. The  $M_{\textup{star}}$ contrasts in the annuli are in  the range 1.1--1.5 (Table~\ref{tab:Mstar-Mgas-3annuli}), as opposed to 1.7--2.5 for the global values (Table~\ref{table:ksTest}), and 
are consistent with those derived by \citet{Meidt+2021}. However, we derive much lower contrasts, by about a factor 3, between arm/interarm medians for $M_{\textup{gas}}$ than what derived by \citet{Meidt+2021}. Part of the difference can be attributed to the smaller spatial scale ($\sim$150~pc as opposed to our $\sim$1~kpc) used by \citet{Meidt+2021}, which reduces spatial averaging. In addition, we only analyze two galaxies, while \citet{Meidt+2021} includes 67 galaxies in their analysis, which increases statistics. 

Annular SFR contrasts between arm and interarm regions are also generally lower than their global counterparts, with the exception of the outermost annuli, which tend to be higher. However, the annular SFR contrasts are non--significant or barely significant, in general.

For sSFRs, the contrasts between arm and interarm regions in the annuli range between 1.3 and 2.3 for the two galaxies, which is slightly larger than the global contrast values, but these differences remain non- or barely significant in all cases. Similar considerations apply to the SFE. Interestingly, we observe trends in SFE contrasts with galactocentric distance, increasing with larger radius in NGC\,628 and decreasing in NGC\,4321. However, as already stated, the annular SFE contrasts are all non- or barely significant.

In summary, with few exceptions, the radial trend analysis indicates that contrasts between  the median values of all parameters considered, $M_{\textup{star}}$  $M_{\textup{gas}}$, SFR, sSFR and SFE, generally decrease  when medians are measured in annuli  rather than across the entire galaxy.

\begin{table}[]
    \centering
    \begin{tabular}{c|c|c|c|c|c|c|c}
    \hline
    Annulus & Parameter & Arm & Interarm &  Arm & Interarm & Median ratio &  KS-test\\
 & & \#spaxels  & \#spaxels & median (log) & median (log) & (Arm/Interarm) & p-value\\
    \hline
    NGC 628 \\
    \hline
        R1 & $M_\textup{star}$ & 55 & 42 & 7.851 & 7.728 & 1.327 & 0.002\\
        R2 & $M_\textup{star}$ & 64 & 194 & 7.250 & 7.165 & 1.217 & 0.271\\
        R3 & $M_\textup{star}$ & 28 & 145 & 6.794 & 6.613 & 1.515 & $5.216\times 10^{-5}$\\
    \hline
    NGC 4321\\
    \hline
    R1 & $M_\textup{star}$ & 18 & 43 & 8.710 & 8.670 & 1.098 & 0.608\\
    R2 & $M_\textup{star}$ & 30 & 127 & 8.276&  8.151 & 1.332 & 0.050\\
    R3 & $M_\textup{star}$ & 18 & 99 & 6.794 & 6.613 & 1.354 & 0.289\\
    \hline
    NGC 628 \\
    \hline
        R1 & SFR$_\textup{10Myr}$ & 55 & 42 & -2.569 & -2.798 & 1.695 & 0.018\\
        R2 & SFR$_\textup{10Myr}$ & 64 & 194 & -2.708 & -3.150 & 2.765 & $5.639\times10^{-5}$\\
        R3 & SFR$_\textup{10Myr}$ & 28 & 145 & -3.020 & -3.597 & 3.777 & 0.004\\
    \hline
    NGC 4321\\
    \hline
    R1 & SFR$_\textup{10Myr}$ & 18 & 43 & -1.611 & -1.877 & 1.844 & 0.103\\
    R2 & SFR$_\textup{10Myr}$ & 30 & 127 & -1.989 & -2.190 & 1.587 & 0.057\\
    R3 & SFR$_\textup{10Myr}$ & 18 & 99 & -2.589 & -3.015 & 2.666 & 0.336\\
    \hline
    NGC 628 \\
    \hline
        R1 & sSFR$_\textup{10Myr}$ & 55 & 42 & -10.387 & -10.519 & 1.354 & 0.358\\
        R2 & sSFR$_\textup{10Myr}$ & 64 & 194 & -9.984 & -10.352 & 2.335 & 0.003\\
        R3 & sSFR$_\textup{10Myr}$ & 28 & 145 & -9.808 & -10.070 & 1.825 & 0.368\\
    \hline
    NGC 4321\\
    \hline
    R1 & sSFR$_\textup{10Myr}$ & 18 & 43 & -10.296 & -10.497 & 1.589 & 0.324\\
    R2 & sSFR$_\textup{10Myr}$ & 30 & 127 & -10.155 & -10.276 & 1.320 & 0.416\\
    R3 & sSFR$_\textup{10Myr}$ & 18 & 99 & -10.470 & -10.724 & 1.796 & 0.538\\
    \hline
    NGC 628 \\
    \hline
        R1 & $M_\textup{gas}$ & 55 & 42 & 7.031 & 6.993 & 1.093 & 0.378\\
        R2 & $M_\textup{gas}$ & 64 & 192 & 6.350 & 6.505 & 0.700 & 0.111\\
        R3 & $M_\textup{gas}$ & 26 & 127 & 6.022 & 6.035 & 0.969 & 0.267\\
    \hline
    NGC 4321\\
    \hline
    R1 & $M_\textup{gas}$ & 18 & 43 & 7.579 & 7.568 & 1.026 & 0.690\\
    R2 & $M_\textup{gas}$ & 30 & 124 & 7.195 & 7.067 & 1.341 & 0.195\\
    R3 & $M_\textup{gas} $& 16 & 92 & 6.375 & 6.255 & 1.320 & 0.520\\
    \hline
    NGC 628 \\
    \hline
        R1 & SFE & 55 & 42 & -9.588 & -9.697 & 1.283 & 0.598\\
        R2 & SFE & 64 & 192 & -9.228 & -9.606 & 2.386 & $3.518\times 10^{-5}$\\
        R3 & SFE & 26 & 127 & -9.098 & -9.514 & 2.605 & 0.022\\
    \hline
    NGC 4321\\
    \hline
    R1 & SFE & 18 & 43 & -9.239 & -9.604 & 2.314 & 0.029\\
    R2 & SFE & 30 & 124 & -9.099 & -9.175 & 1.190 & 0.958\\
    R3 & SFE & 16 & 92 & -9.249 &  -9.162 & 0.818 & 0.742\\
    \hline

    \end{tabular}
   \caption{Medians, median ratios, and KS-tests of the same parameters as Table~\ref{table:ksTest} in three annuli each for NGC~628 and NGC~4321.  
    }
    \label{tab:Mstar-Mgas-3annuli}
\end{table}

\subsection{Sanity Check on SFR Estimates}
As our original goal was to improve on SFR determinations by using SED fits rather than 1-- or 2--band recipes, we test our results against one of those recipes. We calculate the SFRs in the spaxels using the combination of UV+24$\mu$m as indicator, and compare the resulting values against our best-fit SFRs from MAGPHYS. We use the formula from \cite{Liu+2011}:
\begin{equation}
    SFR (\lambda_1, \lambda_2)= C_{\lambda_1}[L(\lambda_1)_{obs} + a_{2} L(\lambda_2)_{obs}].
\end{equation}
where SFR is in unit of $M_\odot$yr$^{-1}$ and the observed luminosities $L(\lambda_1)_{obs}$ and $L(\lambda_2)_{obs}$ are both defined as $\lambda$L($\lambda$) in units of erg s$^{-1}$, with $\lambda_1$=FUV (0.153$\mu$m) and $\lambda_2$=24~$\mu$m. The two constants are $C_{\lambda_1}= 4.6\times 10^{-44}$ and $a_{2}=6.0$ for local star--forming galaxies \citep{Liu+2011, Calzetti2013}. The main assumption for this recipe is that the SFR needs to remain constant for the relevant timescales of the FUV and 24~$\mu$m emission, $\approx$100~Myr \citep{KennicuttEvans2012}. Like all recipes, the accuracy of the SFR derived with the formula above is sensitive to the SFH assumed for the region.

\begin{figure}
    \centering
    \includegraphics[width=0.45\linewidth]{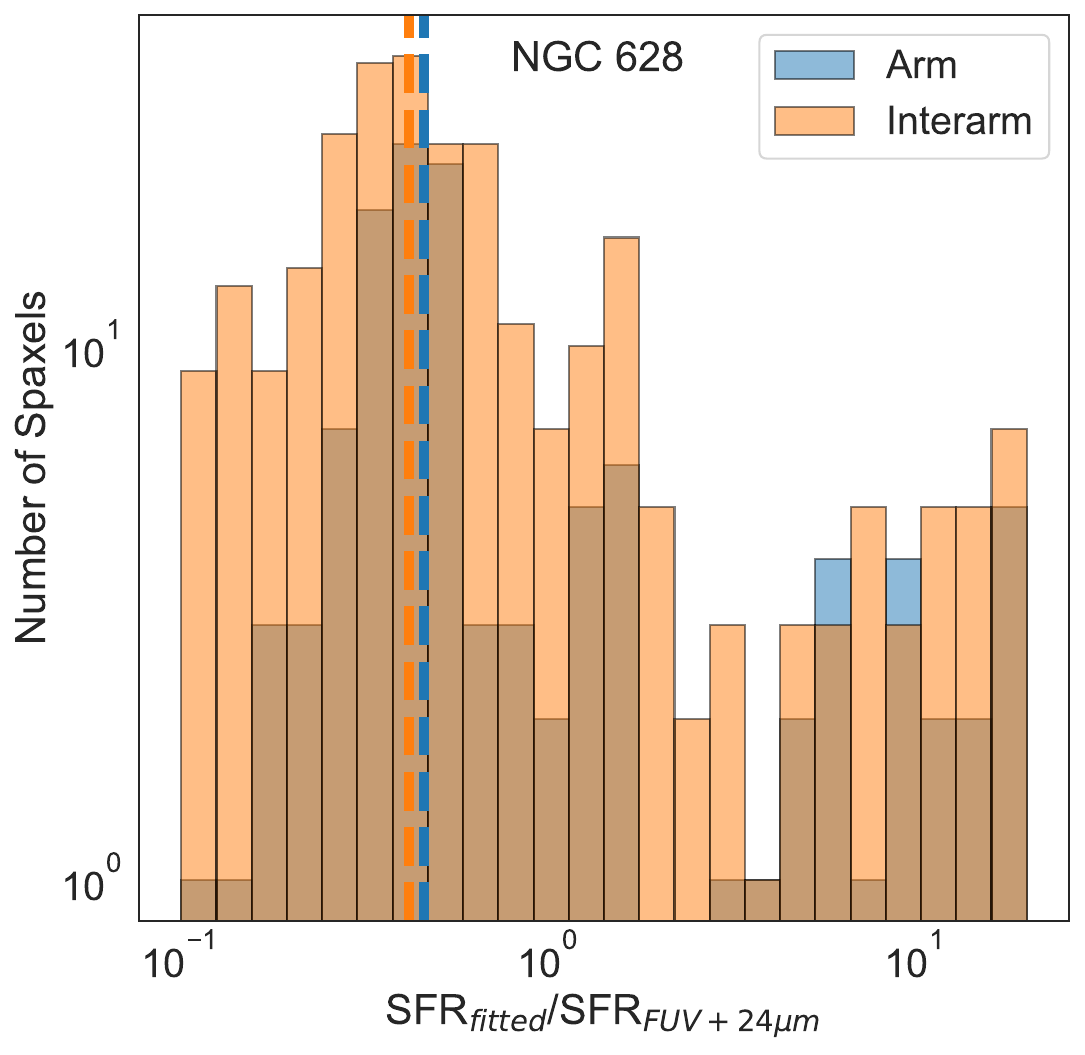}
    \includegraphics[width=0.45\linewidth]{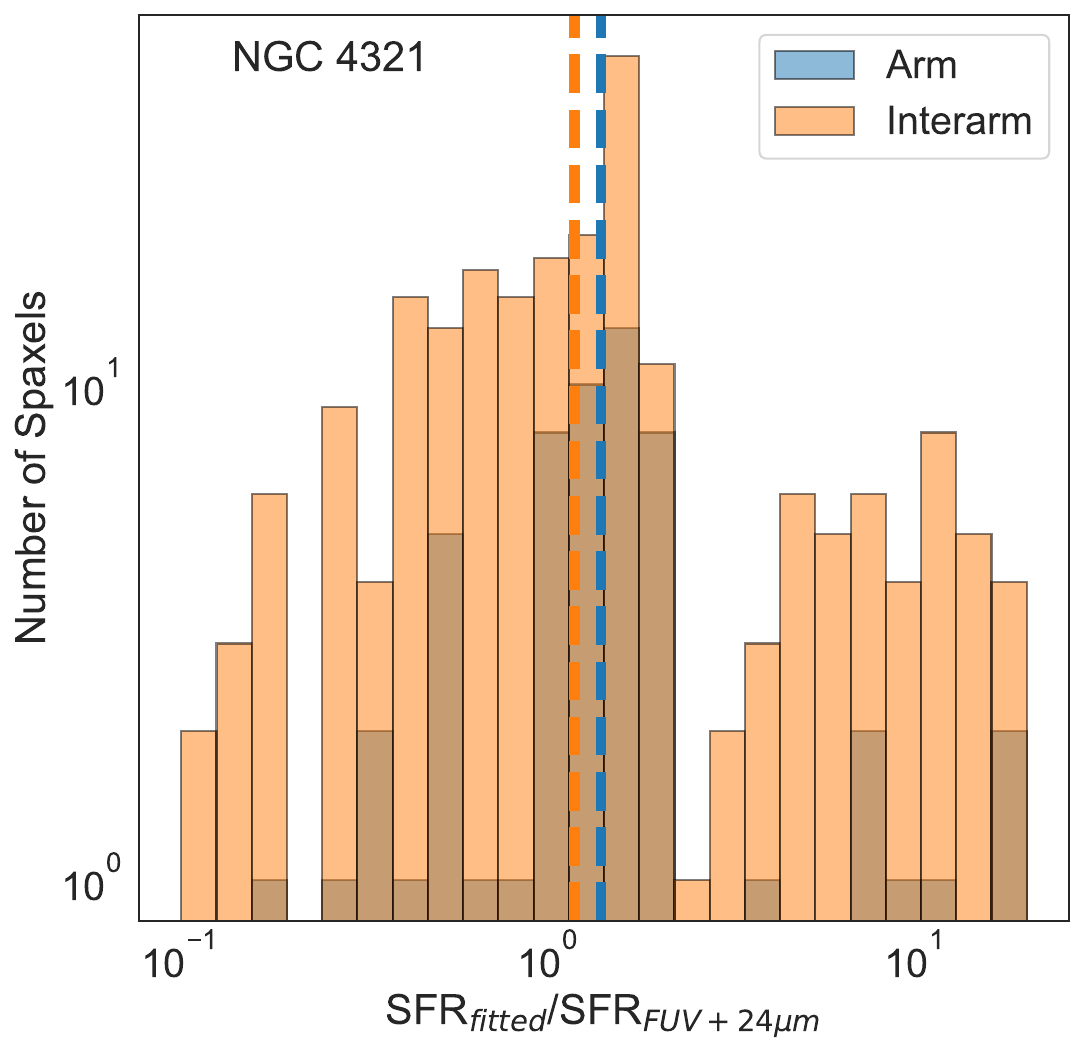}
    \caption{The best-fit SFRs compared with SFRs calculated using UV+24$\mu$m (see text). Note that the vertical scale is logarithmic.  In both panels, the vertical dashed lines mark the median of the distribution. Median of the left panel: 0.458 (arm) and 0.417 (interarm); median of the right panel: 1.386 (arm) and 1.175 (interarm).
    }
    \label{fig:SFR_ratios}
\end{figure}

The results of our comparison are shown in Figure \ref{fig:SFR_ratios}. In both galaxies, we find that the UV+24~$\mu$m recipe yields values for the SFR whose median agrees within a factor $\sim$2 with the SFRs from the best--fit SEDs. However, the spread is large, spanning the range 0.1$\lesssim$SFR$_{fitted}$/SFR$_{FUV+24\mu m}\lesssim$40 and with a notable overdensity for values of SFR$_{fitted}$/SFR$_{FUV+24\mu m}>$3--4. We find that while the large ratios are found in both arms and interarm regions, the smallest ratios are predominantly found in the interarm regions. Based on the fit results, we find that the large variations in SFR ratios are due to the more sophisticated assumptions on the SFH afforded by SED fitting routines. Galactic regions are characterized by a complex relation between SFR and luminosity and better fit by a range of different timescales than the fixed one assumed by the recipes (constant SFR for $\sim$100~Myr for FUV and 24~$\mu$m). For instance, interarm regions are likely to contain sporadic star formation on top of a generally quiescent population, which imply lower SFR than those derived from UV+24~$\mu$m.  

Our summary of the above considerations is that, when $\sim$1~kpc regions are analyzed in galaxies, SED fitting  may provide a more reliable tool to derive SFRs than 1- or 2--band recipes. This is because of the dependence of SFR recipes on the SFH of the region, which lends itself to larger location--dependent variatons than those of whole galaxies. Our considerations assume that energy balance between dust absorption and emission works well at this scales, which is a reasonable assumption down to at least 1~kpc \citep{Smith+2018}.

\subsection{Discussion: Comparison with Previous Work}

Our basic result is that, when using SED fitting to derive physical parameters for regions within galaxies, we don't observe large variations in either sSFR and SFE between arm and interarm regions for the two spirals we have analyzed: NGC~628 and NGC~4321. The variations are at the level of a factor two or lower, and when they are larger, this difference is non--significant ($<$3~$\sigma$), when removing radial trends from the measurements. 

Much of the existing observational work employs recipes to derive SFR and stellar masses, which our independent SED--fitting approach has put to the test. Our results are in better agreement with previous results that find only modest or negligible excess of sSFR and SFE in the arms relative to the interarm regions \citep{Elmegreen+1986, Foyle2010, Kreckel+2016, Foyle+2011,  HuangKauffmann2015, Hart+2017, Querejeta2021, Querejeta+2024}. Studies of our Galaxy have also found negligible variations in SFE \citep{Moore+2012} and cluster formation efficiency \citep{Eden+2013} between spiral arms and interarm regions, indicating a limited role for spiral arms as triggers of star formation \citep{Ragan+2018} and supporting a physical scenario in which star formation is a local process \citep{Urquhart+2021}. 

We find that arm/interarm contrasts for M$_{star}$ are relatively modest in our annulus--based analysis, in agreement with previous, recipe--based results, using, e.g., color--corrected H--band maps \citep{Zibetti+2009} or dust--corrected 3.6~$\mu$m maps \citep{Meidt+2021}. In comparison, SFR contrasts can be as large as factors 3--4 from SED fits, also in agreement with recipe--based derivations \citep[e.g.][]{Foyle2010, Querejeta+2024}. 

Albeit with limited significance, we find that in NGC\,628 the contrasts in SFE are larger or comparable to those in sSFR, while the opposite occurs in NGC\,4321. One proposed `gatherer' model suggests that sSFR contrasts should be larger than SFE contrasts, due to the non--linear response of the gas, and of the SFR,  to perturbations in the gravitational potential \citep{Meidt+2021}. Our results cannot support or refute this model, given the limited size of our sample of two galaxies.

Simulations support a relatively minor role of spiral arms and other structures in star formation. As already mentioned in the Introduction, \citet{Dobbs+2011} and \citet{Kim+2020} reach similar conclusions in regard to the fact that SFRs should be typically enhanced by factors 2 or less in the presence of spiral arms, and that the main role of the arms is to gather material rather than trigger star formation. 
\cite{Tress+2020} simulate interacting galaxies, finding that, while the interactions introduce morphological changes in the spiral arms, the star formation rate (SFR) remains very similar in both interacting and non--interacting galaxies. They conclude that the SFR is primarily governed locally by 
feedback within the cold interstellar medium instead of arm structures. 
Similarly, \cite{Semenov+2017}  conclude from their simulations  that feedback and dynamical processes play the dominant role in the self--regulation and low efficiency of star formation in galaxies.

\subsection{Uncertainties and Limitations}
A major assumption in our analysis is that the stellar Initial Mass Function (IMF) is constant and universal, both between arms and interarm regions, and from galaxy to galaxy. 
While there is still controversy on this issue \citep[e.g.][]{Bastian+2010}, claims of IMF variations \citep[e.g.][]{Cappellari+2012, Lee+2009} have subsequently been shown to be explainable with variations in SFHs and/or with statistical sampling, or be relegated to extreme environmental conditions \citep{Weisz+2012, Andrews+2014, Newman+2017}.
As the regions we investigate are not extreme in density or SFR, and are large enough to avoid issues of small number statistics (mass in stars $\gg 10^6 M_{\odot}$ in each spaxel), we adopt a universal IMF as a reasonable assumption for our study.

A second major assumption is that the energy balance approach of MAGPHYS can be applied over our 1--1.5 kpc regions. This assumption is tested in \cite{Battisti+2019} and found to be reasonable.
The traveling distances of the ionizing photons leaking out of HII regions are of–order a few hundred pc to a kpc within the disks of galaxies \citep{Haffner+2009}.
Furthermore, the dust IR emission is usually contained within the local several hundred parsecs of the heating region \citep{Lawton+2010}.
Both lines of evidence suggest that regions $\gtrsim$1 kpc can be considered energetically independent, enabling us to adopt energy balance between the stellar light absorbed by dust in the UV–NIR and the dust emission in the MIR–FIR within each region.

\section{Summary and Conclusions}
We have investigated the star formation properties of the arm and interarm regions of two nearby spiral galaxies, NGC 628 and NGC 4321. 
We have divided each galaxy into a few hundreds spaxels with  $\sim$1--1.5 kpc size and constructed the multi–wavelength SEDs of each region, with each SED containing about 20 -- 22 broad--band photometric datapoints from the UV to the FIR. The SEDs probe both the stellar and dust emission from each spaxel. We have used the SED--fitting code MAGPHYS to model the SED of each spaxel, and extract the best--fit  SFRs (averaged over 10~Myr) and the stellar masses at each spaxel location, as well the age of the luminosity--weighted oldest stars. We have also included public CO maps to construct molecular gas masses for the spaxels. 
We have applied the publicly available masks to separate the arm and interarm regions for both galaxies. We present the distributions of the sSFRs, the ages of the oldest stars, and the SFEs, separately  for  arms and interarm regions, both globally and as a function of galactocentric distance.

We find that the constrasts  between arm and interarm regions, as measured from the ratio of the medians, for sSFR, SFE and the ages of the oldest stars, in both NGC~628 and NGC 4321, are typically within a factor of 2. The differences in the distributions are generally not statistically significant. 

Taken at face value, the results for these two galaxies provide strong support for the scenario that spiral arms are the ``gatherers'' of material instead of acting as the ``triggers'' of star-formation. In the case of the sSFR, one would expect an increase in the SFR relative to the background stellar mass if spiral arms were triggers of star  formation. However, we show that {\em both} SFRs and stellar mass are higher in the arm regions than in the interarm region. This is supporting evidence that the spiral arms simply `collect' material, be it stars, as traced by the stellar mass distribution, or the gas from which new stars can form, as traced by the SFR distribution. If confirmed by additional analyses, this scenario supports simulation results that star formation is a local, cloud--level physical process, and not a global process influenced by galaxy--wide triggers.

The two galaxies in the present pilot study represent too small of a sample for any generalization of our conclusions, and some ambiguities are still present in our results. For instance, while contrasts in the sSFR and SFE between arm and interarm regions are generally low, there are variations even within the same galaxy. For instance, in NGC~628, the contrasts in the SFE vary between 1.3 and 2.6 as a function of galactocentric distance, although the significance of these variations is low.
In the near future, we plan on extending our  analysis to the entire sample of 34 spiral galaxies from the combination of the SINGS and KINGFISH surveys to provide firmer tests of the two scenarios for the role of spiral arms in star formation.

\section*{Acknowledgement}
The authors thank the anonymous referee whose detailed comments have greatly improved the paper.

This research has made use of the NASA/IPAC Extragalactic Database (NED) which is operated by the Jet Propulsion Laboratory, California Institute of Technology, under contract with the National Aeronautics and Space
Administration.

B.S. and D.C. acknowledge support from the grant NASA ADAP/2021, ID 80NSSC22K0478, `1,000 star  formation histories in nearby galaxies' for this research.\\

\software{
astropy \citep{2013A&A...558A..33A},
numpy \citep{harris2020},
matplotlib \citep{matplotlib},
scipy \citep{2020SciPy-NMeth},
photutils \citep{photutils},
SWarp \citep{Bertin1996},  
MAGPHYS \citep{daCunha2008MNRAS}
Seaborn \citep{Waskom2021}
}

\appendix

\section{The Marginalized Likelihood Probability Functions of the best fit parameters}\label{AppendixA}
\begin{figure}%
    \centering
    \includegraphics[width=8.5cm]{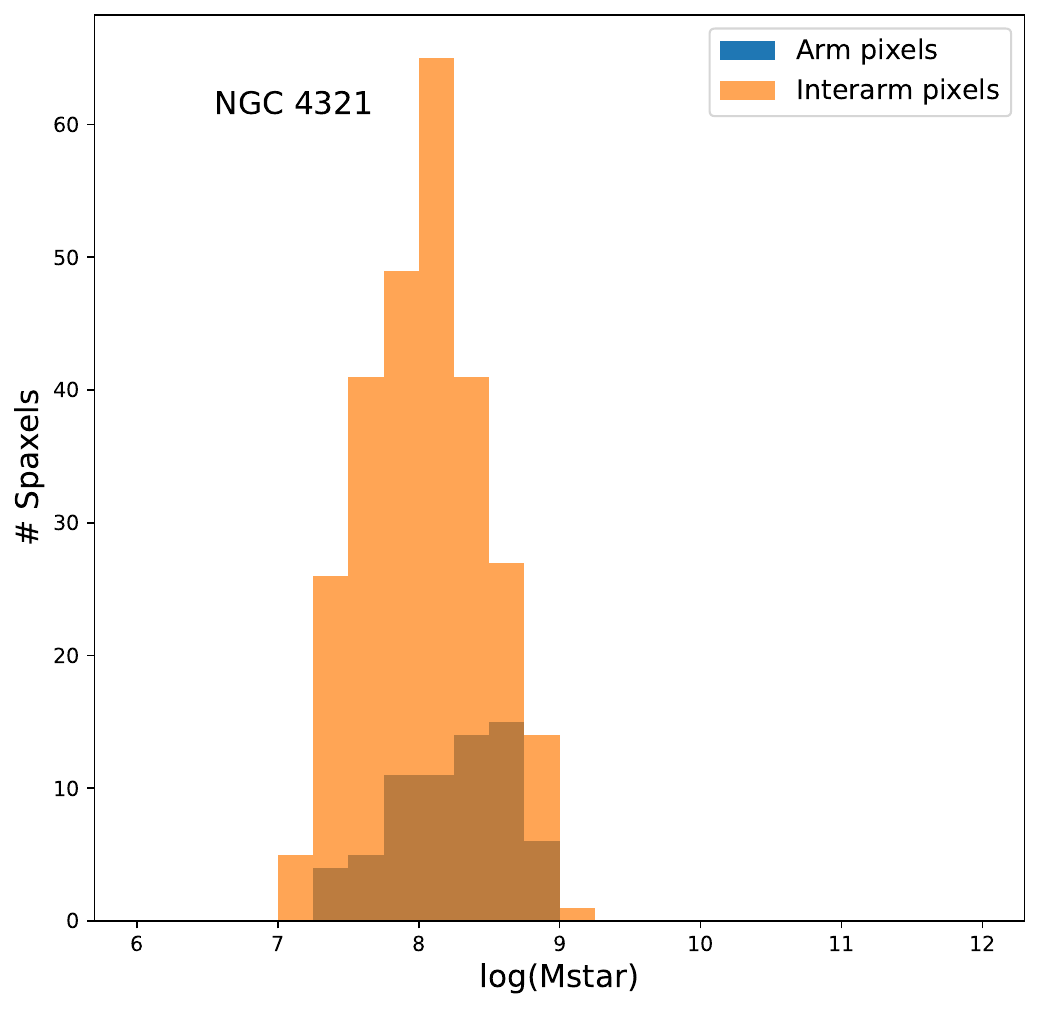} 
    \includegraphics[width=8.5cm]{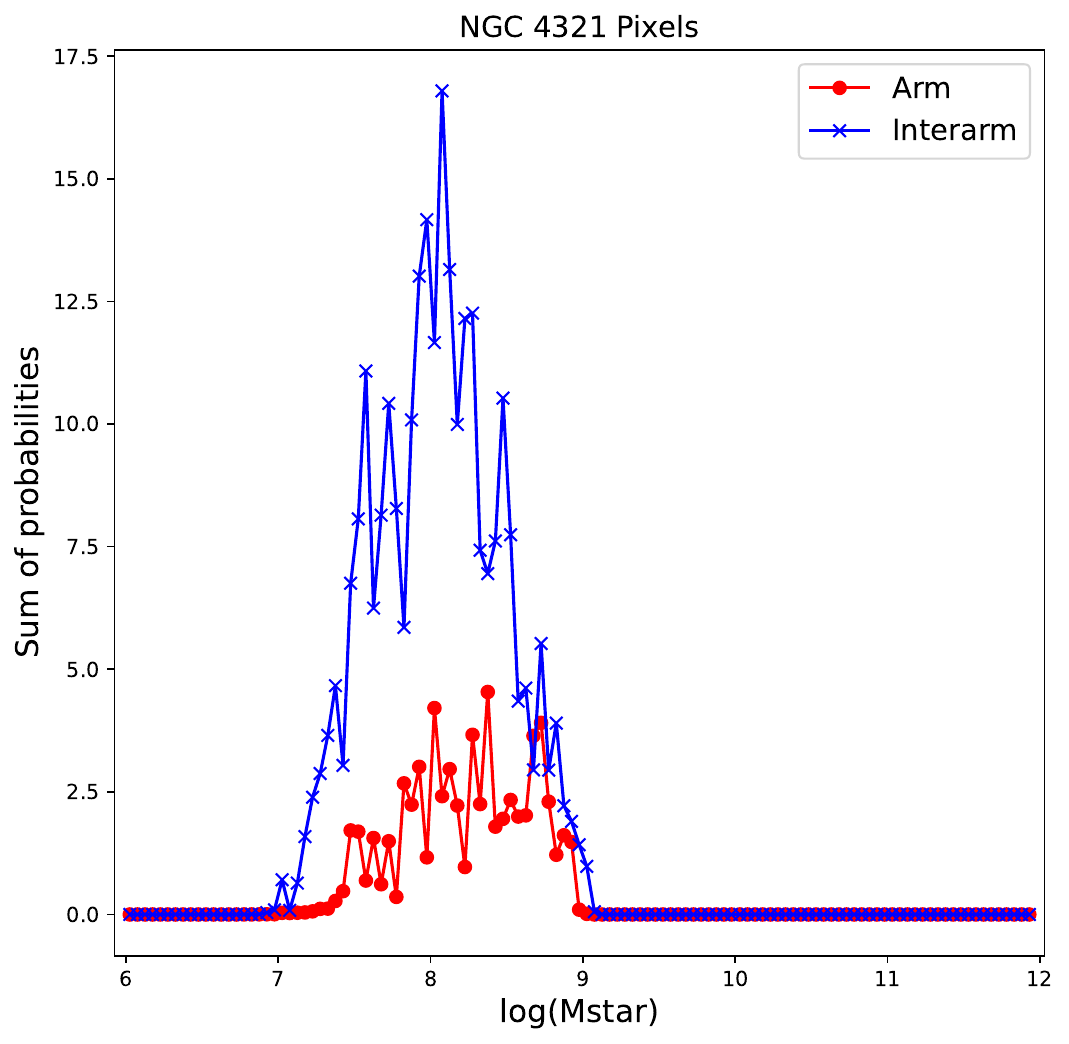} 
    \includegraphics[width=8.5cm]{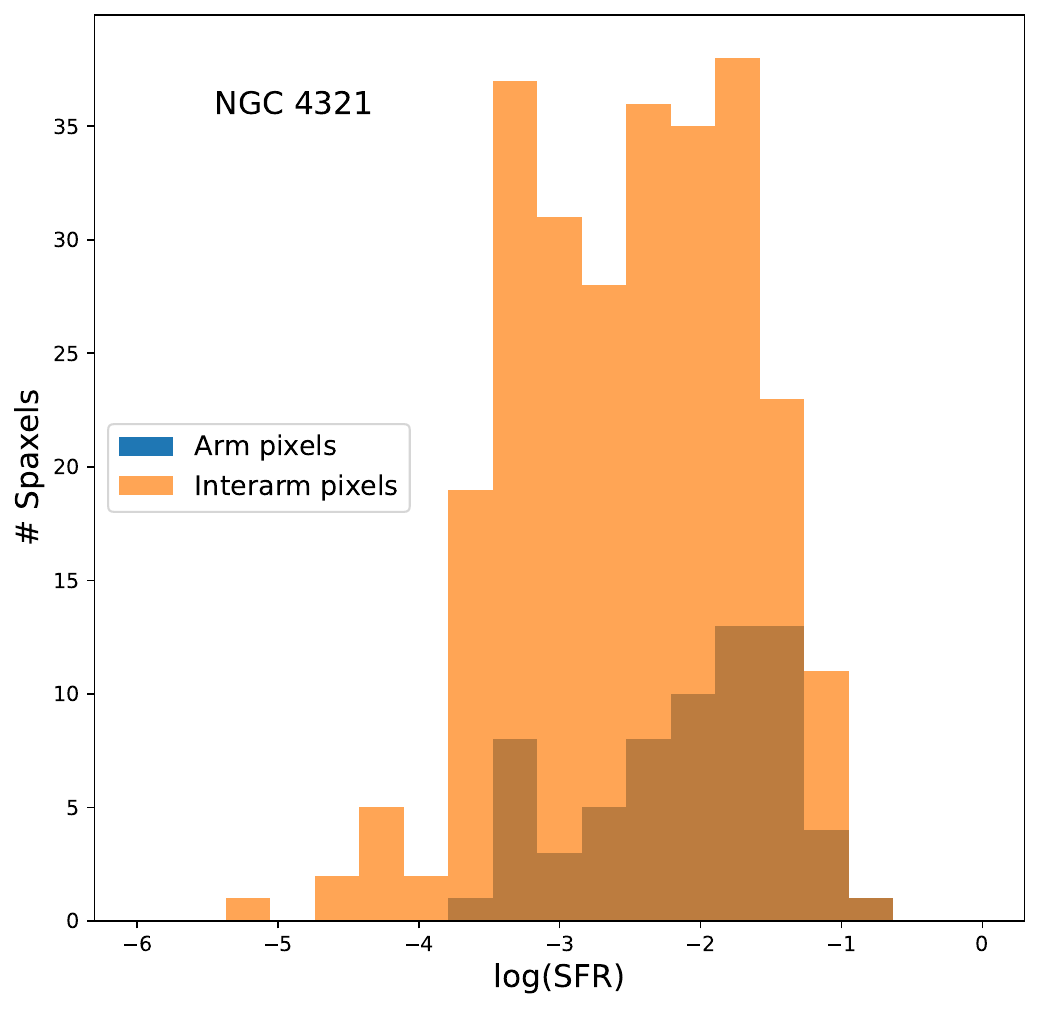} 
    \includegraphics[width=8.5cm]{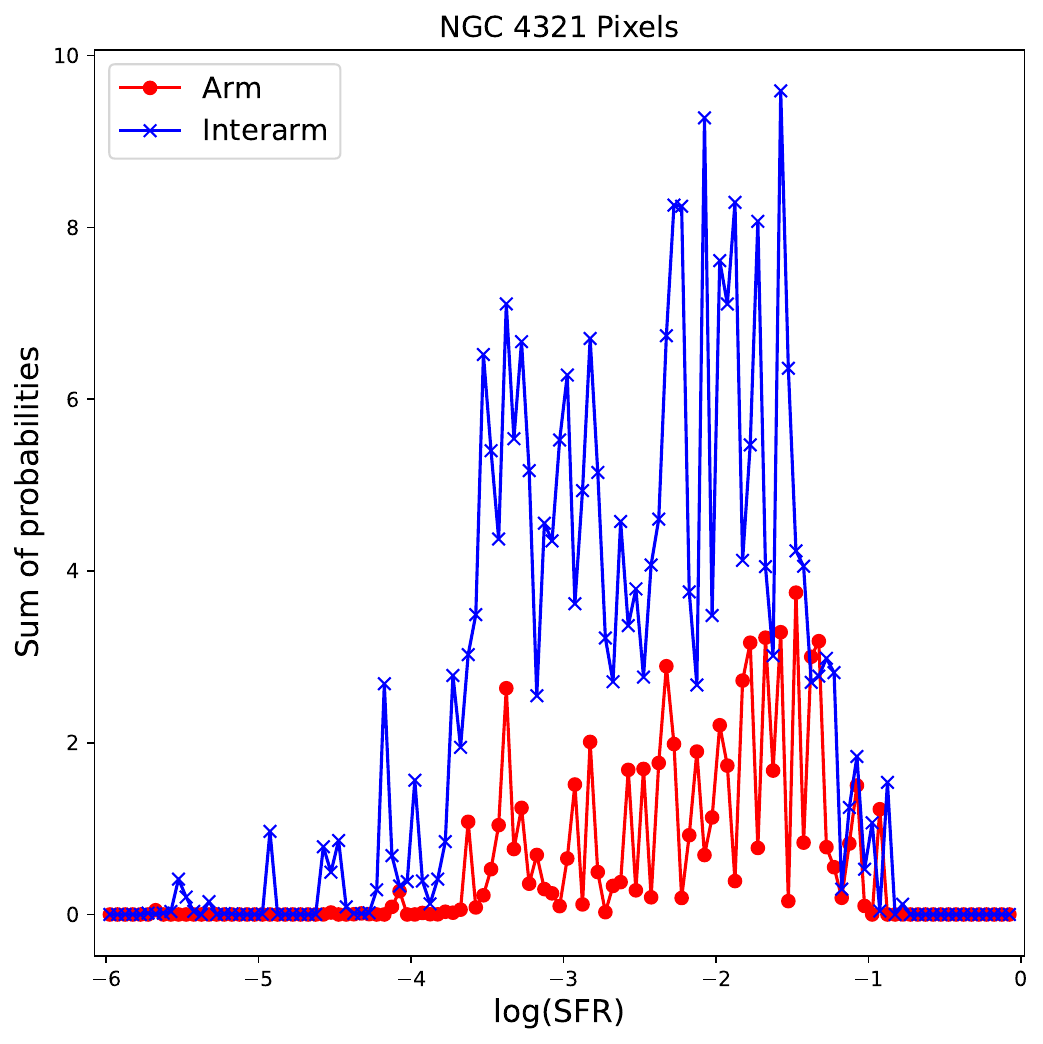} 
    \caption{The distribution of best-fit stellar masses of arm/interarm spaxels (regions) in NGC 4321. Upper left panel: spaxel number counts v.s. best-fit values. Upper right panel: sum of all the likelihood distributions output by MAGPHYS for arm/interarm spaxel.
    The lower row: same as the upper row but for SFRs. } 
    \label{fig:appendixPDFs}%
\end{figure}

We conclude that there are no significant differences for either distribution (sSFR, SFE, and  $t_{\text{form}}$) between arm and interarm regions, in both NGC 628
and NGC 4321. This conclusion comes from the analysis of the distributions of the best-fit values of these two parameters. However, the MAGPHYS code  returns likelihood distributions, i.e., the probability distribution functions (PDFs) of each parameter rather than a single best-fit value, so a legitimate question is whether the uncertainties on the best-fit parameter values affect the conclusions. In this appendix, we try to address this possibility.\\
In our SED fitting, the width of all the parameter PDFs are very narrow, due to the wide wavelength coverage and the high S/N of our data. 
We give two examples of the sum of best-fit parameter PDFs for M$_{star}$ and SFR in Figure \ref{fig:appendixPDFs} . In the left panels of both rows, we show the distribution of the best-fit parameter values of the arm and interarm spaxels in NGC 4321. In the right panels of each row, we show the sum of all the pixel PDFs for each parameter. As we can see, adding the PDFs does not change the overall shape of the distribution, but leads to noisier distributions. Thus, changing the approach (from histograms of best fit values to sums of probability distributions) does not change our results and conclusions, but only adds noise. 

\section{Tests of spiral arm cuts}\label{AppendixB}
In section \ref{sec:method}, we overlay the \cite{Querejeta2021}'s masks on our images and attribute spaxels to arm regions if more than 50\% of their area fall within the arms as defined by the masks. Here we show the results of two stricter cuts, to test whether the 50\% choice affects our conclusions. We use NGC 628 as an example.

\begin{figure}
    \centering
    \includegraphics[width=8.5cm]{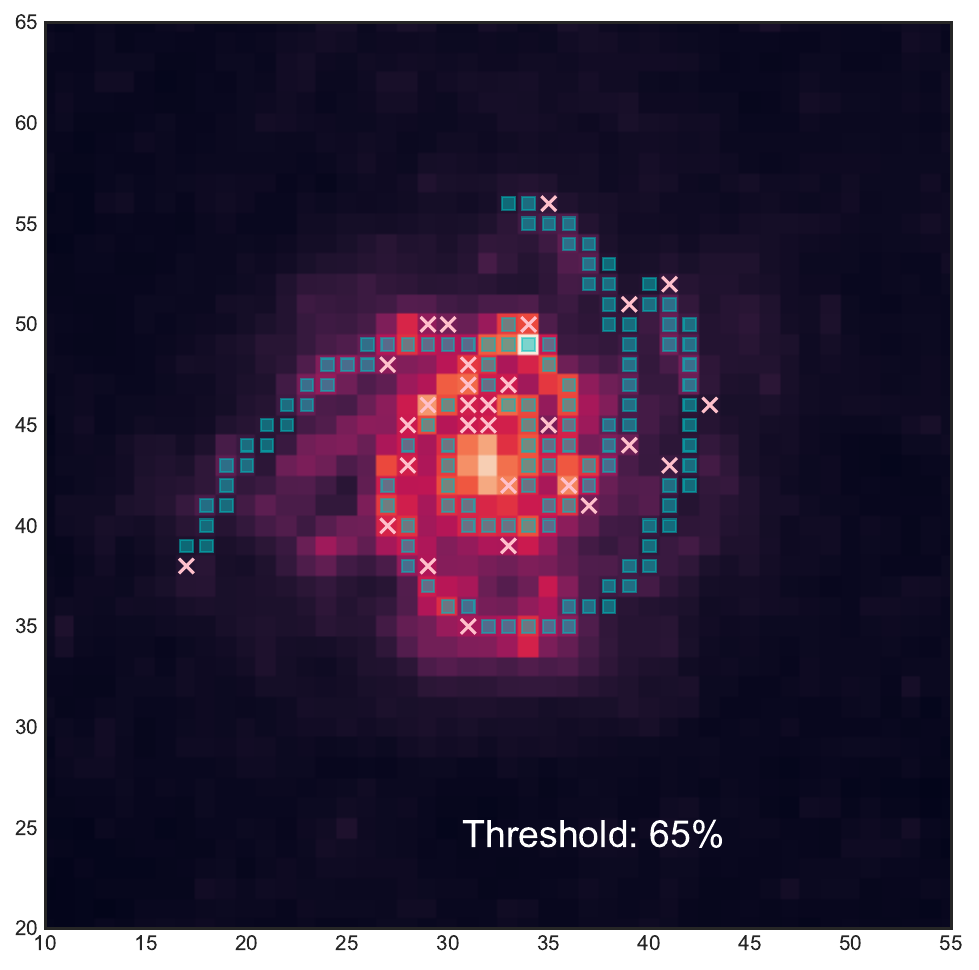} 
    \includegraphics[width=8.5cm]{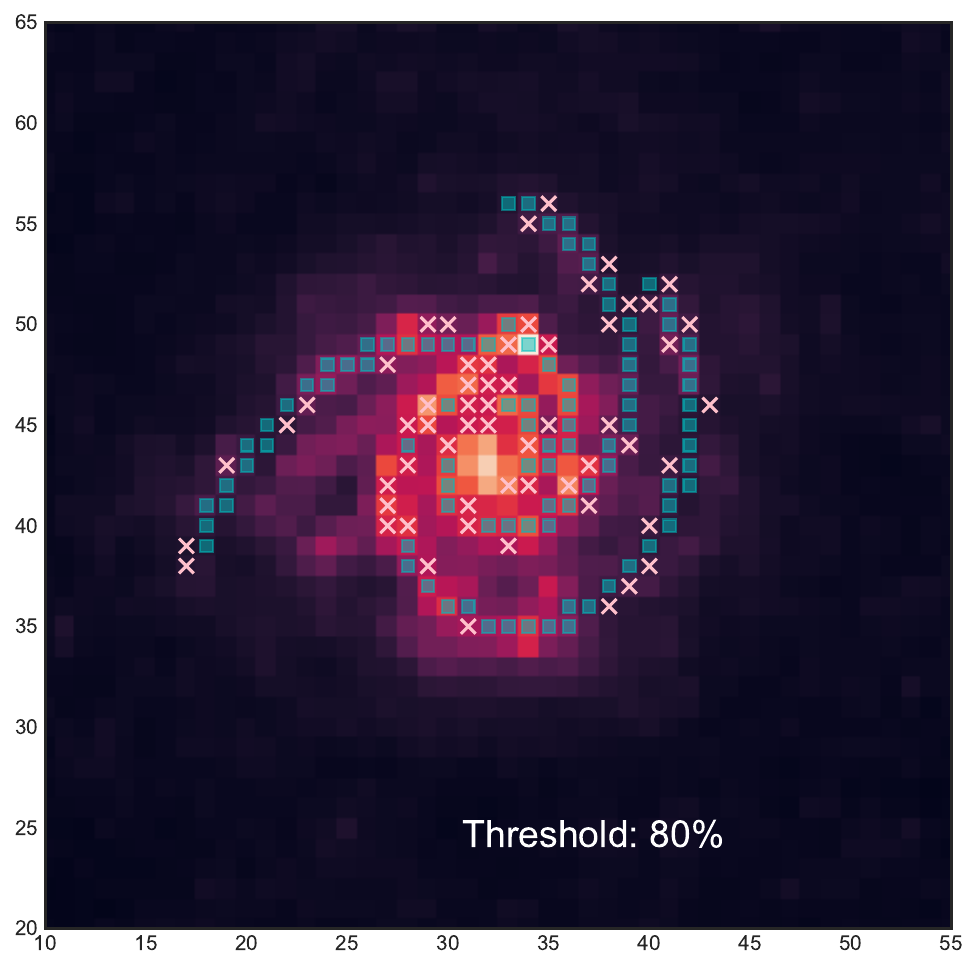} 
    \caption{Spaxel locations of arm regions under new cut thresholds. IN both panels, the pink cross marks the spaxels we excluded comparing with the 50\% threshold in the paper. These spaxels are marked as neither arm nor interarm in the present analysis. }
    \label{fig:spaxel_loc_two_cuts}%
\end{figure}

The left panel of Figure \ref{fig:spaxel_loc_two_cuts} shows the location of spaxels that have 65\% or more  of their area that falls within the arms defined by the masks are marked as arm regions. The right panel shows the arm spaxels under the condition that 80\% or more of their area falls within the mask for the spiral arms. In both cases, a sufficiently large number of spaxels remain to perform an analysis. 

\begin{figure}
    \centering
    \includegraphics[width=0.8\linewidth]{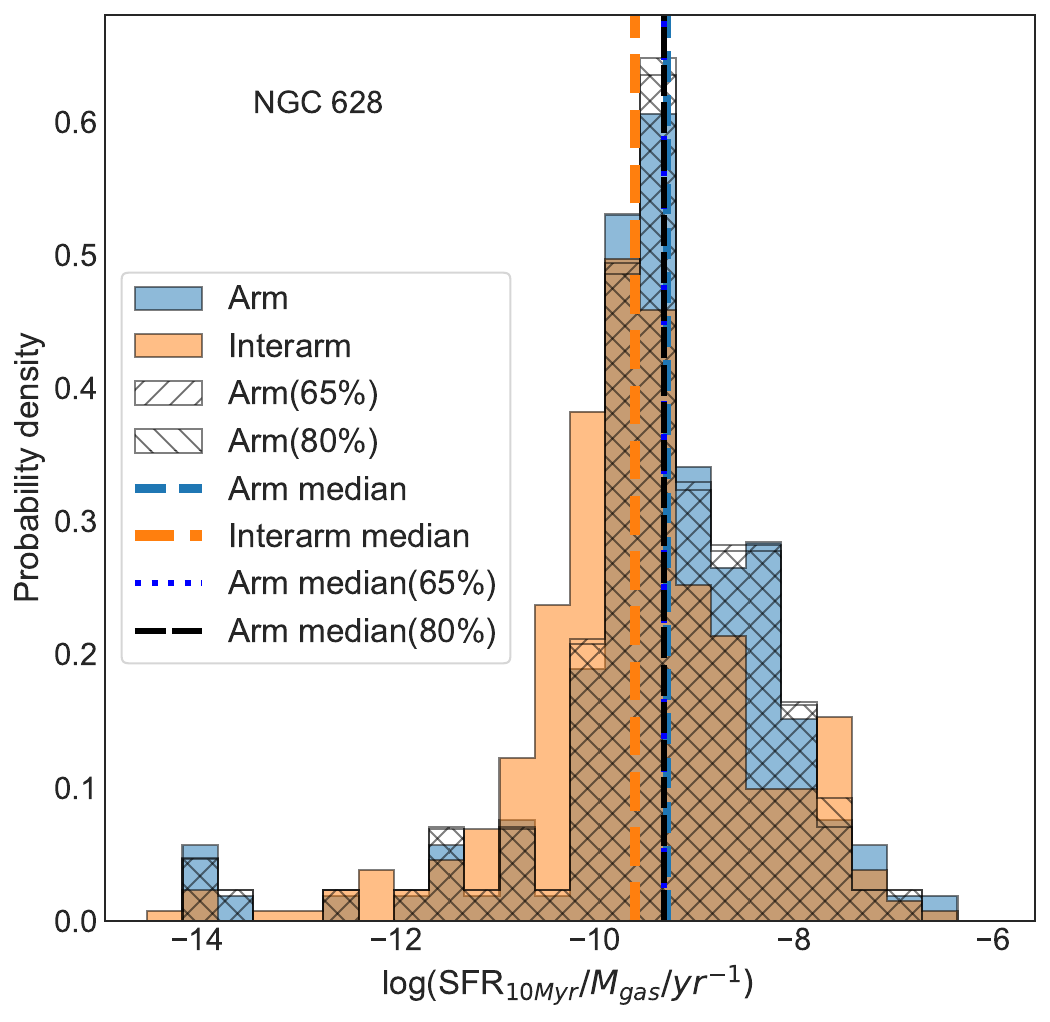} 
     \caption{The distribution of SFEs for the arm and interarm spaxels of NGC~628. For the arm regions, the three cases of 50\%, 65\% and 80\% of arm areal coverage for a spaxel to be attributed to the spiral arms are shown.}
\label{fig:SFE_three_thresholds}%
\end{figure}

Figure~\ref{fig:SFE_three_thresholds} shows the distribution of SFEs for the arm and interarm regions of NGC~628. For the arm regions, the three cases that 50\%, 65\% and 80\%  of the area of arm spaxels falls within the spiral arm mask are shown. The three distribution look very similar to each other, with minimal differences. In all three cases, the differences in the SFE medians are negligible, indicating that our choice of a 50\% threshold for arm spaxels does not affect our conclusions.

\bibliography{references}{}
\bibliographystyle{aasjournal}

\end{document}